\documentclass[
aps,
pre,
preprint,
superscriptaddress,
amsmath,amssymb
]{revtex4-2}
\usepackage{dcolumn}

\usepackage{amssymb}
\usepackage{hyperref}
 \hypersetup{
    colorlinks = true,
    citecolor = blue,
    linkcolor = red,
    }

\usepackage{amsthm}
\usepackage{graphicx}
\usepackage{epsfig}

\usepackage{scalerel}
\usepackage{dblfloatfix}% http://ctan.org/pkg/dblfloatfix
\usepackage{enumitem}
\usepackage[utf8]{inputenc}
\usepackage{xcolor, soul}
\sethlcolor{yellow}

\usepackage{tikz}
\usepackage{physics}

\newcommand{\ii}{\mathrm{i}\,}
\begin{document}

\preprint{APS/123-QED}
\title{High-frequency tuning of internal resonance and targeted energy transfer in a Van der Pol oscillator coupled to a nonlinear energy sink}

\author{Somnath Roy}
\email{roysomnath63@gmail.com}
\affiliation{Institute of Engineering \& Management, University of Engineering \& Management, Kolkata 700091, India}
    ```
\author{Mattia Coccolo}
\email{mattiatommaso.coccolo@urjc.es}
\affiliation{Nonlinear Dynamics, Chaos and Complex Systems Group, Departamento de Geología, Física y Química Inorgánica, Universidad Rey Juan Carlos, Tulipán s/n, 28933 Móstoles, Madrid, Spain}

\author{Sayan Gupta}
\email{sayan@iitm.ac.in}
\affiliation{Department of Applied Mechanics and Biomedical Engineering, IIT Madras, Chennai 600036, Tamil Nadu, India}
\affiliation{Centre for Complex Systems and Dynamics, IIT Madras, Chennai 600036, Tamil Nadu, India}

\author{Miguel A.F. Sanjuán}
\email{miguel.sanjuan@urjc.es}
\affiliation{Nonlinear Dynamics, Chaos and Complex Systems Group, Departamento de Geología, Física y Química Inorgánica, Universidad Rey Juan Carlos, Tulipán s/n, 28933 Móstoles, Madrid, Spain}
\affiliation{Royal Academy of Sciences of Spain, Valverde 22, 28004 Madrid, Spain}

\begin{abstract}

\noindent
Targeted energy transfer (TET) from a Van der Pol oscillator coupled to a nonlinear energy sink (NES) is investigated under the action of a high-frequency external drive, which tunes the effective natural stiffness and promotes resonance capture, facilitating energy transfer. Using \textit{direct partition of motion} with \textit{complexification averaging}, the mechanism of energy flow and instability control through \textit{hopf bifurcation} is characterized. A spectrally evaluated Q-factor, based on FFT at the effective slow frequency, captures the resonance peaks indicating the efficient energy transfer. Finally, the energy-dissipation metric is consistent with these Q-maps and identifies the regions where transient energy pumping is most effective.

\vspace{0.5cm}

\noindent\textbf{Keywords:} Nonlinear energy sink; Targeted energy transfer; Van der Pol oscillator; Internal resonance; Vibrational resonance; Hopf bifurcation

\end{abstract}

\maketitle

\section{Introduction}
The mechanism of flow of energy in nonlinear coupled vibrating systems is a central challenge in both dynamical and acoustical systems, as it has important connotations in a host of applications, which can be broadly categorized into vibration suppression, such as isolation \cite{lu2013investigation} and absorption \cite{starosvetsky2009vibration}, and vibration utilization, specifically energy harvesting \cite{yang2021nonlinear}.
The objective in vibration-based energy harvesting applications is to maximize the conversion of 
mechanical energy into electrical energy.  

While there exists a host of active control strategies by which vibration isolation and/or vibration absorption can be enhanced, energy harvesting systems predominantly rely on the passive mechanisms where mechanical energy is harvested  \cite{balaji2021applications}. The widely used method in this category includes attaching secondary dynamical systems to the main structure, which act as an energy absorber.  Maximizing vibration energy dissipation through the secondary structure is the main goal by tuning the frequencies of the coupled systems in a suitable ratio (super or sub-harmonics). In this context, the use of geometric nonlinearities to effectively tune the natural frequencies in relation to the excitation spectrum has been the subject of a large body of research, which aims to improve vibration isolation, vibration absorption, and energy harvesting mechanisms \cite{sun2015multi,liu2019new,yang2016modeling,yan2023design}.\\

In vibration-based energy harvesting involving coupled systems, the objective is to enhance the energy transfer to a secondary attachment, where mechanical energy is typically converted into electrical energy through a piezoelectric element, while minimizing backscattering by exploiting geometric nonlinearities. A central concept in this context is \textit{targeted energy transfer} (TET) to a \textit{nonlinear energy sink} (NES), namely, a lightweight attachment designed to irreversibly absorb and dissipate broadband vibrational energy~\cite{vakakis2009nonlinear,vakakis2022nonlinear}.
The energy exchange is enabled by \textit{internal resonance}, which arises when one or more modes of the primary system satisfy a commensurability condition with the secondary system (NES), such as $m\omega_1=n\omega_2$~\cite{asadi2018nonlinear,chen2023internal}. Under these conditions, resonance capture through \textit{nonlinear normal modes}~\cite{vakakis1997non,avramov2013review} gives rise to complex dynamical phenomena, including amplitude modulation, beating, and sustained energy exchange. This passive mechanism has been extensively studied in both single- and multi-degree-of-freedom systems, including configurations with ungrounded~\cite{lee2005complicated,das2023nonlinear} and grounded~\cite{gendelman2001energy} NES attachments, as well as in nonsmooth vibro-impact settings~\cite{nucera2007targeted}. Beyond passive operation, active control by external periodic forcing at commensurate frequencies has also been widely investigated. It is well known that, when the forcing amplitude exceeds a critical threshold, the system may enter a \emph{strongly modulated response} regime characterized by a continuous, quasiperiodic, and nearly irreversible transfer of vibrational energy from the primary oscillator to the NES~\cite{gendelman2006quasiperiodic}. Under sustained excitation, the primary system may exhibit \emph{amplitude saturation}, a characteristic signature of TET, whereby the vibration amplitude remains nearly constant despite further increases in the forcing amplitude. 
This effect suppresses large oscillations and provides robust vibration mitigation~\cite{gourdon2007nonlinear}. More generally, TET enables substantial vibration reduction in the primary system through the attachment of a properly tuned NES. In this regime, a significant fraction of the energy associated with large-amplitude oscillations of the primary system is transferred to the NES and dissipated, leading to a marked reduction of the primary response~\cite{gendelman2010bifurcations,wang2023vibration}.
\\

Despite the considerable body of work on passive and actively forced TET, here we propose a different active-control strategy based on the application of an external forcing at a frequency much higher than the characteristic frequencies of the system. This high-frequency drive acts on the primary oscillator and, through the system nonlinearities, effectively renormalizes the system parameters, thereby generating an averaged or effective dynamics. In this way, the resonance condition can be actively controlled, allowing the targeted energy transfer process to be tuned. This mechanism may be viewed as an extension of the well-known phenomenon of \emph{vibrational resonance} (VR), in which the response to a weak low-frequency signal is amplified by increasing the amplitude of a high-frequency drive~\cite{landa2000vibrational,zaikin2002vibrational}. The theoretical basis of vibrational mechanics is the method of \emph{direct partition of motion} introduced by Blekhman~\cite{blekhman2000vibrational}. Since then, numerous applications have been reported, including mechanical oscillators~\cite{jeevarathinam2011theory,fahsi2009suppression}, parametric systems~\cite{roy2021vibrational,roy2023controlling}, neuronal systems~\cite{yu2011vibrational}, plasmas~\cite{roy2016analysis}, nonsmooth oscillators~\cite{roy2025vibrational}, energy harvesting~\cite{coccolo2014energy}, and, more recently, quantum systems~\cite{roy2025controlling} and oscillator arrays~\cite{roy2026collective}. For a comprehensive overview, see the recent review in Ref.~\cite{yang2024vibrational}. Although VR has been studied predominantly in single-degree-of-freedom systems, its role in coupled systems exhibiting internal resonance remains largely unexplored. It is important to distinguish these two mechanisms. In VR, an external high-frequency excitation exploits the intrinsic nonlinearity of the system to enhance the response to a low-frequency signal. By contrast, internal resonance arises from an intrinsic commensurability among the frequencies of coupled modes, whether linear or nonlinear. Introducing a fast external forcing into a coupled system therefore offers the possibility of controlling the internal-resonance condition itself, leading to what may be termed \emph{internal vibrational resonance}~\cite{DjomoMbong2025PLA_IVR}. In that work, the authors considered an autonomous Van der Pol--Duffing--Van der Pol triad. The present study is independent of, and complementary to, that framework. Here we consider a coupled Van der Pol--NES system subjected to an external high-frequency drive applied to the Van der Pol oscillator. We show that this forcing induces an effective 1{:}1 internal resonance while preserving the essential mechanism of vibrational resonance. In addition, we provide a signal-level diagnostic that demonstrates how the vibrational-resonance framework can be implemented numerically in this class of systems. These results open a route toward the active tuning of energy transfer from the primary system to the NES, thereby providing a mechanism for either suppressing or enhancing TET in vibration mitigation and energy-harvesting applications under different operating conditions.
\\

In this article, we consider a coupled system consisting of a Van der Pol oscillator and an essentially nonlinear energy sink (NES). The Van der Pol oscillator is a classical example of a self-sustained oscillator exhibiting stable limit-cycle dynamics, and it can serve as an operational model for micro- and nanoelectromechanical systems (MEMS/NEMS). As such, it provides a relevant framework for investigating energy transfer, stable operation, and robust signal amplification~\cite{asadi2018nonlinear,yousuf2023phononic}. As a canonical self-sustained system with a stable limit cycle, the Van der Pol oscillator, when coupled to a NES, furnishes a minimal setting for the study of targeted energy transfer. In addition, it is a prototypical model for excitable dynamics and, under suitable modifications, can reproduce the behavior of the FitzHugh--Nagumo neuron model; see, e.g., Ref.~\cite{SieweSiewe2015}. This connection suggests that the detection and mapping tools developed here could, in principle, be extended to excitable systems as well.

The application of VR modifies the effective dynamics of the Van der Pol--Duffing--NES system, thereby allowing the internal-resonance condition to be shifted or induced. Furthermore, it enables either enhancement or suppression of the oscillatory response. In this sense, VR constitutes an effective active-control mechanism for adaptive energy harvesting and vibration mitigation.

\section{Mathematical Model}\label{model}

We consider two coupled nonlinear oscillators, where the primary oscillator is driven by an external periodic excitation whose frequency is much higher than the linearized natural frequency of the primary system. The equations of motion are given by:

\begin{figure}[htp]
    \centering
    \includegraphics[scale=0.65]{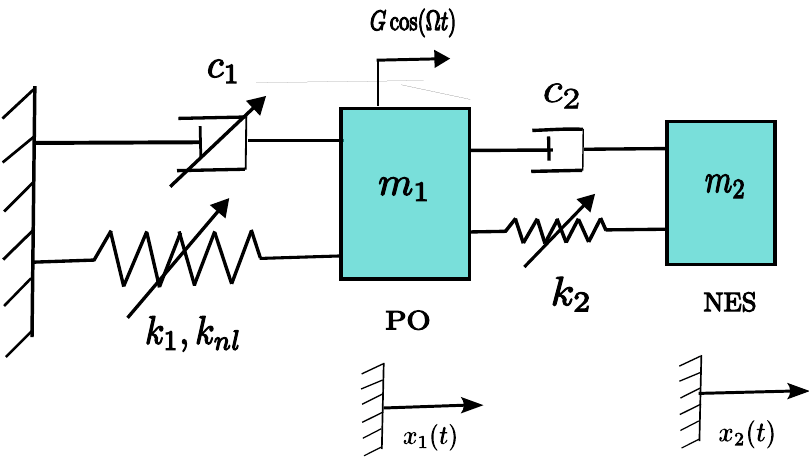}
    \caption{Schematic of the coupled Van der Pol oscillator--NES system under high-frequency excitation of the primary oscillator.}
    \label{fig:schematic}
\end{figure}

\begin{eqnarray}
        &m_1\ddot{x_1} +c_1(x_1^2 - 1)\dot{x_1} + k_1 x_1 + k_{nl} x_1^3 + c_2(\dot{x_1} - \dot{x_2})+ k_2(x_1 - x_2)^3 = G \cos(\Omega t),
        \label{eq:original_PO}\\
        &m_2\ddot{x_2} + c_2(\dot{x_2} - \dot{x_1}) + k_2(x_2 - x_1)^3=0.
    \label{eq:original_NES}
\end{eqnarray}

Equations~\eqref{eq:original_PO} and \eqref{eq:original_NES} describe a Van der Pol oscillator with cubic stiffness nonlinearity coupled to an essentially nonlinear NES; in particular, the secondary system has no linear stiffness term. A schematic representation is shown in Fig.~\ref{fig:schematic}. The generalized displacement variables $x_1$ and $x_2$ denote the vibratory coordinates of the primary oscillator and the NES, respectively. The corresponding masses are $m_1$ and $m_2$. The parameters $k_1$ and $k_{nl}$ represent the linear and nonlinear cubic stiffness coefficients of the primary oscillator, while $c_1$ denotes the nonlinear damping coefficient associated with the Van der Pol element. The parameter $c_2$ characterizes the viscous coupling between the primary oscillator and the NES, and $k_2$ is the nonlinear coupling stiffness between them. The external high-frequency forcing term $G\cos(\Omega t)$ is applied to the primary oscillator and separates the response into fast and slow time scales. These fast oscillations interact with the system nonlinearities and generate effective stiffness and damping contributions that scale with the drive intensity $G/\Omega^2$. As a result, the effective natural frequencies can be tuned in real time so as to satisfy a $1{:}1$ internal-resonance condition, thereby facilitating TET without modifying the intrinsic system parameters. The term $c_1(x_1^2-1)\dot{x}_1$ represents Van der Pol-type nonlinear damping and is responsible for the self-excited oscillations that lead to a stable limit cycle. The cubic terms $k_{nl}x_1^3$ and $k_2(x_1-x_2)^3$ describe Duffing-type nonlinear restoring forces, which are essential for capturing the effects of internal resonance through the interaction of nonlinear normal modes. Finally, the dissipative coupling terms $c_2(\dot{x}_1-\dot{x}_2)$ and $c_2(\dot{x}_2-\dot{x}_1)$ govern the energy exchange between the two oscillatory components.
\\

This model can exhibit \textit{internal resonance}, whereby energy is efficiently exchanged between coupled oscillatory components owing to commensurability among their characteristic frequencies, including primary, subharmonic, and superharmonic resonance conditions. The presence of nonlinear stiffness and damping gives rise to a broad spectrum of dynamical responses, including periodic, quasiperiodic, transient, and chaotic oscillations~\cite{vakakis2009nonlinear}. In the following, we analyze the dynamics of the system under high-frequency excitation, with particular emphasis on the onset of the $1{:}1$ internal-resonance condition, for which the effective natural frequencies of the primary oscillator and the NES become nearly equal. In this regime, the nonlinear coupling is strongest, enabling efficient TET so that energy injected into the primary oscillator is preferentially transferred to the NES and dissipated there.

Equations \eqref{eq:original_PO} and \eqref{eq:original_NES} can be non-dimensionalized to the form:

\begin{eqnarray}
        &\ddot{x_1} +\gamma_1(x_1^2 - 1)\dot{x_1} + \omega
        _0^2 x_1 + \alpha_1 x_1^3 +\gamma_2(\dot{x_1} - \dot{x_2})+\alpha_2(x_1 - x_2)^3 = g \cos(\Omega t),
        \label{eq:modified_PO}\\
        &\mu\ddot{x_2} + \gamma_2(\dot{x_2} - \dot{x_1}) + \alpha_2(x_2 - x_1)^3=0,
    \label{eq:modified_NES}
\end{eqnarray}

\noindent where \(\mu = m_2/m_1\), \(\gamma_1 = c_1/m_1\), \(\omega_0^2 = k_1/m_1\), \(\alpha_1 = k_{nl}/m_1\), \(\gamma_2 = c_2/m_1\), and \(\alpha_2 = k_2/m_1\). The parameter \(g = G/m_1\) is not assumed to be of order \(\mathcal{O}(\epsilon)\), but is instead determined by the magnitude of \(G\). Following the structure of Blekhman's perturbation method, it is treated solely as a high-frequency excitation acting on the primary oscillator in Eq.~\eqref{eq:modified_PO}. The distinct role of the high-frequency input amplitude in the resonance response, which should not be confused with the forcing amplitude \(G\), will be clarified in a later section.

It is mathematically convenient to express Eqs.~(\ref{eq:modified_PO}--\ref{eq:modified_NES}) in terms of the new variables $u=x_1$ and $v=x_1-x_2$, which represent the relative displacements, since the coupling and damping forces depend only on the displacement and velocity differences between the primary oscillator and the NES. Thus Eqs.~\eqref{eq:modified_PO} and \eqref{eq:modified_NES} can now be written as
\begin{eqnarray}
        &\ddot{u} +\gamma_1(u^2 - 1)\dot{u} + \omega
        _0^2 u + \alpha_1 u^3 +\gamma_2\dot{v}+\alpha_2v^3 =  g \cos(\Omega t),
        \label{eq:relative_PO}\\
        &\mu\ddot{v} + \gamma_2\dot{v} + \alpha_2 v^3=\mu\ddot{u}.
    \label{eq:relative_NES}
\end{eqnarray}

\noindent Equations~\eqref{eq:relative_PO} and \eqref{eq:relative_NES}, written in terms of the non-dimensional parameters, are now ready to be analyzed within the framework of \textit{direct partition of motion}, as described in \cite{blekhman2000vibrational}. This approach allows us to characterize the effective dynamics of the coupled primary-oscillator--NES system.

\subsection{Effective dynamics and vibrational resonance}
The high-frequency drive introduces widely separated time scales into the system. The effective slow-scale behavior of the coupled system describes the envelope dynamics and enables deducing the resonance conditions and tuning mechanisms for NES design. The fast dynamics can be averaged out using the Blekhman perturbation method. The technique is a direct partition of motion, which separates motion into a slow and a fast scale. Since multiple coordinate systems are introduced in the analysis, the following table is provided.
\[
\boxed{
\begin{array}{ccl}
\text{Original coordinates} & : & (x_1,\, x_2) \\[4pt]
\Downarrow & & \\[-2pt]
\text{Relative form} & : & (u,\, v), \quad u = x_1, \; v = x_1 - x_2 \\[4pt]
\Downarrow & & \\[-2pt]
\text{Slow–fast decomposition} & : & 
(u,\, v) = (X_1 + \psi_1,\, X_2 + \psi_2)
\end{array}
}
\]

\noindent The dependence on the slow and fast time scales can be written explicitly as

\begin{equation}
\begin{split}
u&=X_1(\omega_0 t,t)+\psi_1(\Omega t, t),\\
v&=X_2(\omega_0 t,t,X_1)+\psi_2(\Omega t, t,\psi_1),
\end{split}
\label{eq:slow_fast}
\end{equation}
where $X_1(t)$ and $X_2(t)$ denote the slowly varying envelopes associated with the slow-scale motion, while $\psi_1(t)$ and $\psi_2(t)$ represent the rapidly oscillating components induced by the high-frequency excitation. Substituting Eq.~\eqref{eq:slow_fast} into Eqs.~\eqref{eq:relative_PO}--\eqref{eq:relative_NES}, the equations of motion governing $X_1$ and $X_2$ can be derived as follows:

\begin{equation}
    \begin{split}
        \ddot{X_1}&+\gamma_1(X_1^2-K)\dot{X_1}+\omega_{p}^2X_1+\gamma_2\dot{X}_2+\alpha_1 X_1^3+\omega_{n}^2 X_2+\alpha_2X_2^3=0,
    \end{split}
    \label{eq:effective_dyn_PO}
\end{equation}

\begin{equation}
    \begin{split}
        \mu\ddot{X}_2+\gamma_2\dot{X}_2+\omega_{n}^2 X_2+\alpha_2X_2^3=\mu\ddot{X}_1
    \end{split}
    \label{eq:effective_dyn_NES}
\end{equation}
These equations describe how the high-frequency drive modifies the effective stiffness and damping of the coupled primary-oscillator--NES system once the fast oscillations have been averaged out. They also capture the slow envelope dynamics on which internal resonance and targeted energy transfer occur, thereby providing a compact description directly relevant to the underlying control mechanisms. The modified (effective) natural frequency of the primary Van der Pol oscillator is given by

\begin{equation}
\omega_{p}=\sqrt{\omega_0^2+\dfrac{3\alpha_1g^2}{2\Omega^2}\left(\dfrac{1}{\Omega^2+\gamma_2^2}\right)}.
\label{eq:effetive_po_freq}
\end{equation}
The detailed derivations of the effective frequencies $\omega_p$, $\omega_n$, and the modified damping parameter $K$ follow from the averaging, and the modified damping parameter, K, follow from the averaging procedure and are presented in Appendix~\ref{appendix:A}. As is evident from Eq.~\eqref{eq:effetive_po_freq}, the natural stiffness of the primary oscillator is modified through the interplay between the nonlinear stiffness $\alpha_1$ and the high-frequency drive strength $g/\Omega^2$. Although no linear stiffness term is present in the original dynamics of the NES, see Eq.~\eqref{eq:original_NES}, an effective linear stiffness emerges in the reduced equation of motion of the NES as a result of the combined influence of the drive strength and the nonlinear coupling $\alpha_2$. This effective stiffness can be identified with the coefficient of the linear term $X_2$ in Eq.~\eqref{eq:effective_dyn_NES} and is given by

 \begin{equation}
     \omega_{n}=\sqrt{\dfrac{3\alpha_2\mu^2g^2}{2\Omega^2}\left(\dfrac{1}{\mu^2\Omega^2+\gamma_2^2}\right)}.
     \label{eq:effetive_nes_freq}
 \end{equation}
Another key observation arises from the modification of the Van der Pol damping nonlinearity by the additional term
\begin{equation}
    K=1-\dfrac{g^2}{2\Omega^4},
    \label{eq:effetive_damping}
\end{equation} 
which modifies the stability characteristics of the primary oscillator through the control parameter $g/\Omega^2$. Furthermore, Eqs.~\eqref{eq:effetive_po_freq} and \eqref{eq:effetive_nes_freq} reveal the possibility of achieving a $1:1$ resonance condition, ($\omega_{p}=\omega_{n}$), between the primary oscillator and the NES by appropriately tuning the amplitude of the high-frequency forcing, $g/\Omega^2$. Assuming $\Omega>>\gamma_2$, this resonance condition can be expressed as follows:

\begin{equation}
    \left(\dfrac{g}{\Omega^2}\right)_{res}\approx\omega_0\sqrt{\dfrac{2}{3(\alpha_2-\alpha_1)}}.
    \label{eq:resonance_fast_amp}
\end{equation}
Equation~\eqref{eq:resonance_fast_amp} identifies the internal resonance peak at which the response amplitude attains its maximum. In the limit of large fast frequency $\Omega$, namely when $\left(\mathcal{O}(\mu\Omega)>>\gamma_2\right)$, this resonance condition becomes independent of the mass ratio $\mu$. However, when the effect of the dissipative coupling $\gamma_2$ is retained, both $\omega_p$ and $\omega_n$ depend on $g$ through different rational functions of $(\Omega^{2}+\gamma_2^{2})$ and $(\mu^{2}\Omega^{2}+\gamma_2^{2})$, respectively. As a consequence, the condition $\omega_p = \omega_n$ leads to a higher-order equation for $\left(g/\Omega^{2}\right)_{\mathrm{res}}$, which is solved numerically. Equation~\eqref{eq:resonance_fast_amp} thus establishes the fundamental connection between the resonance condition and the strength of the high-frequency signal, which constitutes one of the central themes of this work. The stability of the primary oscillator oscillations can be tuned primarily via the effective damping parameter $K$ (equivalently, through $g/\Omega^2$) in Eq.~\eqref{eq:effetive_damping}. It is well-known from Eq.~\eqref{eq:effective_dyn_PO} that the nonlinearity in damping is responsible for the transition from the sustained limit cycle oscillation to oscillation death through a \textit{Hopf bifurcation}, depending on the sign of $K$. Thus, the critical bifurcation point can be obtained by setting $K=0$, which yields

\begin{equation}
    g_{\mathrm{hopf}}=\sqrt{2}\Omega^2.
    \label{eq:hopf_point}
\end{equation}

Thus, when $K > 0$ (i.e., $g < g_{\mathrm{hopf}}$), a limit-cycle oscillation can be observed, whereas for $K < 0$ (i.e., $g > g_{\mathrm{hopf}}$), the system undergoes suppressed oscillatory motion leading to oscillation death. To validate all the analytically predicted results, numerical simulations have been performed in Sec.~\eqref{sec:numerical}.

\subsection{Slow flow of effective dynamics}

To describe the slow-time evolution of the amplitudes and phases of the coupled modes, we derive the slow-flow equations for the effective dynamics by introducing the complex variables

\begin{eqnarray}
    Z_1=\dot{X_1}+\ii\omega X_1,\\
    \label{eq:complex_Z1}
    Z_2=\dot{X_2}+\ii\omega X_2. \label{eq:complex_Z2}
\end{eqnarray}
The variables $Z_1$ and $Z_2$ serve to represent the dynamics dictated by Eq.~\eqref{eq:effective_dyn_PO} and \eqref{eq:effective_dyn_NES} in a rotating complex frame, thus facilitating the transformation of displacement and velocity in a single complex term. The variables $X_j, j=1,2$ can be expressed,

\begin{eqnarray}
    X_j=-\dfrac{\ii}{2\omega}\left(Z_j-Z^*_j\right),\label{eq:complex_X1}\\
    \dot{X}_j=\dfrac{1}{2}\left(Z_j+Z^*_j\right),\label{eq:complex_X1_dot}
\end{eqnarray}

\noindent where $Z^*$ represents the complex conjugate of $Z$, $j=1,2$ and $\mathrm{i}=\sqrt{-1}$. By substituting Eqs.~\eqref{eq:complex_X1} and \eqref{eq:complex_X1_dot} into Eqs.~\eqref{eq:effective_dyn_PO} and \eqref{eq:effective_dyn_NES}, the governing equations can be written in complex form as

\begin{equation}
\begin{split}
    \dot{Z}_1&-\ii\omega Z_1-\dfrac{\gamma_1}{8\omega^2}(Z_1-Z_1^*)^2(Z_1+Z_1^*)\\
    &-\dfrac{\gamma_1K}{2}(Z_1+Z_1^*)+\dfrac{\ii\alpha_1}{8\omega^3}(Z_1-Z_1^*)^3+\dfrac{\gamma_2}{2}(Z_2+Z_2^*)-\dfrac{\ii\omega}{2}(Z_2-Z_2^*)+\dfrac{\ii\alpha_2}{8\omega^3}(Z_2-Z_2^*)^3=0\label{eq:complex_PO}
\end{split}
\end{equation}
and,

\begin{equation}
    \mu\left(\dot{Z}_2-\dfrac{\ii\omega}{2}(Z_2+Z_2^*)\right)-\dfrac{\ii\omega}{2}\left(Z_2-Z_2^*\right)+\dfrac{\gamma_2}{2}(Z_2+Z_2^*)+\dfrac{\ii\alpha_2}{8\omega^3}(Z_2-Z_2^*)^3=\mu\left(\dot{Z}_1-\dfrac{\ii\omega}{2}(Z_1+Z_1^*)\right)\label{eq:complex_NES}
\end{equation}

\noindent  Equations~\eqref{eq:complex_PO} and~\eqref{eq:complex_NES} govern the evolution of the complex slow variables $Z_1$ and $Z_2$, which encode both the amplitude and phase of the effective dynamics. This complex representation provides a compact description of the coupled system, in which the nonlinearities and coupling terms appear as polynomial interactions involving $Z_j$ and their complex conjugates $Z_j^*$. To characterize the slow amplitude modulation more explicitly, we introduce the following transformation in terms of a slowly varying complex amplitude $\phi$ and a fast oscillation at frequency $\omega$:

\[
Z_1 = \phi_1(t) e^{\ii \omega t}, 
\qquad 
Z_2 = \phi_2(t) e^{\ii \omega t}.
\] 
Equations~\eqref{eq:complex_PO} and \eqref{eq:complex_NES} can then be expressed in terms of $\phi_1$ and $\phi_2$ as:

\begin{equation}
    \begin{split}
       \dot{\phi}_1e^{\ii\omega t}&-\dfrac{\gamma_1}{8\omega^2}\left(\phi_1^3e^{3\ii\omega t}-\mid\phi_1\mid^2\phi_1e^{\ii\omega t}-\mid\phi_1\mid^2\phi_1^*e^{-\ii\omega t}+{(\phi_1^*)}^3e^{3\ii\omega t}\right)-\dfrac{\gamma_1K}{2}\left(\phi_1e^{\ii\omega t}+\phi_1^*e^{-\ii\omega t}\right)\\
       &+\dfrac{\ii\alpha_1}{8\omega^3}\left(\phi_1^3e^{3\ii\omega t}-3\mid\phi_1\mid^2\phi_1e^{\ii\omega t}+3\mid\phi_1\mid^2\phi_1^*e^{-\ii\omega t}+{(\phi_1^*)}^3e^{3\ii\omega t}\right)-\dfrac{\ii\omega}{2}\left(\phi_2e^{\ii\omega t}-\phi_2^*e^{-\ii\omega t}\right)\\
       &+\dfrac{\gamma_2}{2}\left(\phi_2e^{\ii\omega t}+\phi_2^*e^{-\ii\omega t}\right)+\dfrac{\ii\alpha_2}{8\omega^3}\left(\phi_2^3e^{3\ii\omega t}-3\mid\phi_2\mid^2\phi_2e^{\ii\omega t}+3\mid\phi_2\mid^2\phi_2^*e^{-\ii\omega t}+{(\phi_2^*)}^3e^{3\ii\omega t}\right)=0
    \end{split}
\end{equation}

and

\begin{equation}
    \begin{split}
    &\mu\left(\dot{\phi}_2e^{\ii\omega t}-\dfrac{\ii\omega}{2}\phi_2^*e^{-\ii\omega t}\right)-\left(1-\epsilon\right)\dfrac{\ii\omega}{2}\phi_2e^{\ii\omega t}+\dfrac{\gamma_2}{2}\left(\phi_2e^{\ii\omega t}+\phi_2^*e^{-\ii\omega t}\right)\\
    &+\dfrac{\ii\alpha_2}{8\omega^3}\left(\phi_2^3e^{3\ii\omega t}-3\mid\phi_2\mid^2\phi_2e^{\ii\omega t}+3\mid\phi_2\mid^2\phi_2^*e^{-\ii\omega t}+{(\phi_2^*)}^3e^{3\ii\omega t}\right)=\mu\left(\dot{\phi}_1e^{\ii\omega t}+\dfrac{\ii\omega}{2}\phi_1e^{\ii\omega t}-\dfrac{\ii\omega}{2}\phi_1^*e^{-\ii\omega t}\right)
    \end{split}
\end{equation}

Using the complexification--averaging approach described in Ref.~\cite{vakakis2003dynamics}, we impose the $1{:}1$ resonance condition by assuming $\omega\approx\omega_p\approx \omega_n$. The slow flow is then obtained by averaging over the fast time scale $\omega t$, i.e., over the carrier oscillations at frequency  $\omega$, which are much faster than the slow envelope variations of $X_1$ and $X_2$. This procedure yields the effective slow-envelope dynamics in terms of the complex amplitudes $\dot{\phi}_1$ and $\dot{\phi}_2$, governed by the following coupled differential equations

\begin{equation}
    \begin{split}
        \dot{\phi}_1&+\dfrac{\gamma_1}{8\omega_p^2}\mid\phi_1\mid^2\phi_1-\dfrac{\gamma_1K}{2}\phi_1\\
        &-\ii\dfrac{3\alpha_1}{8\omega_p^3}\mid\phi_1\mid^2\phi_1+\dfrac{\gamma_2}{2}\phi_2-\ii\dfrac{\omega_p}{2}\phi_2-\ii\dfrac{3\alpha_2}{8\omega_p^3}\mid\phi_2\mid^2\phi_2=0,
    \end{split}
\end{equation}

\begin{equation}
    \begin{split}
        \mu\dot{\phi}_2-(1-\epsilon)\dfrac{\ii\omega_p}{2}\phi_2+\dfrac{\gamma_2}{2}\phi_2-\ii\dfrac{3\alpha_2}{8\omega_p^3}\mid\phi_2\mid^2\phi_2=\mu\left(\dot{\phi}_1+\dfrac{\ii\omega_p}{2}\phi_1\right).
    \end{split}
\end{equation}

Introducing the polar representations
\[
\phi_1 = A e^{\mathrm{i}\theta_1}, \qquad \phi_2 = B e^{\mathrm{i}\theta_2},
\]
where $A$ and $B$ are real amplitudes, and $\theta_1$ and $\theta_2$ are real phases, the slow-flow equations for the amplitudes and phases are obtained by separating the real and imaginary parts and setting them equal to zero. This yields

\begin{equation}
    \begin{split}
        \dot{A}=-\dfrac{\gamma_1}{8\omega_p^2}A^3+\dfrac{\gamma_1K}{2}A-\dfrac{\gamma_2}{2}B\cos(\theta_2-\theta_1)-\dfrac{\omega_p}{2}B\sin(\theta_2-\theta_1)-\dfrac{3\alpha_2}{8\omega_p^3}B^3\sin(\theta_2-\theta_1)
    \end{split}
    \label{eq:A_dynamics}
\end{equation}

\begin{equation}
    \begin{split}
       A\dot{\theta}_1=\dfrac{3\alpha_1}{8\omega_p^3}A^3-\dfrac{\gamma_2}{2}B\sin(\theta_2-\theta_1)+\dfrac{\omega_p}{2}B\cos(\theta_2-\theta_1)+\dfrac{3\alpha_2}{8\omega_p^3}B^3\cos(\theta_2-\theta_1)
    \end{split}
    \label{eq:theta1_dynamics}
\end{equation}
and,
\begin{equation}
    \begin{split}
       \mu\dot{B}=-\dfrac{\gamma_2}{2}B+\epsilon\dot{A}\cos(\theta_1-\theta_2)-\epsilon A\sin(\theta_1-\theta_2)\dot{\theta}_1-\dfrac{\mu\omega_p}{2}A\sin(\theta_1-\theta_2)
    \end{split}
    \label{eq:B_dynamics}
\end{equation}

\begin{equation}
    \begin{split}
        \mu B\dot{\theta}_2=\dfrac{(1-\epsilon)\omega_p}{2}B+\dfrac{3\alpha_2}{8\omega_p^3}B^3+\epsilon\dot{A}\sin(\theta_1-\theta_2)+\epsilon A\cos(\theta_1-\theta_2)\dot{\theta}_1+\dfrac{\mu\omega_p}{2}A\cos(\theta_1-\theta_2)
    \end{split}
    \label{eq:theta2_dynamics}
\end{equation}

\noindent 

The evolution of the slow amplitudes $A$ and $B$ provides qualitative insight into the stability of the oscillations and its tuning by the high-frequency drive. The amplitude $A$ characterizes the effective response of the primary Van der Pol oscillator, whereas $B$ describes the envelope dynamics of the NES. Linearization of Eq.~\eqref{eq:A_dynamics} shows that, for $K>0$, the primary oscillator possesses a stable limit cycle with a finite steady-state amplitude. As the system approaches the critical point $K=0$, the amplitude $A$ collapses and the oscillation vanishes, indicating a supercritical Hopf bifurcation. In the following section, we present numerical results and discuss these predictions.

\begin{figure}[htp]
\begin{center}
\includegraphics[width=15.5cm, ,clip=true]{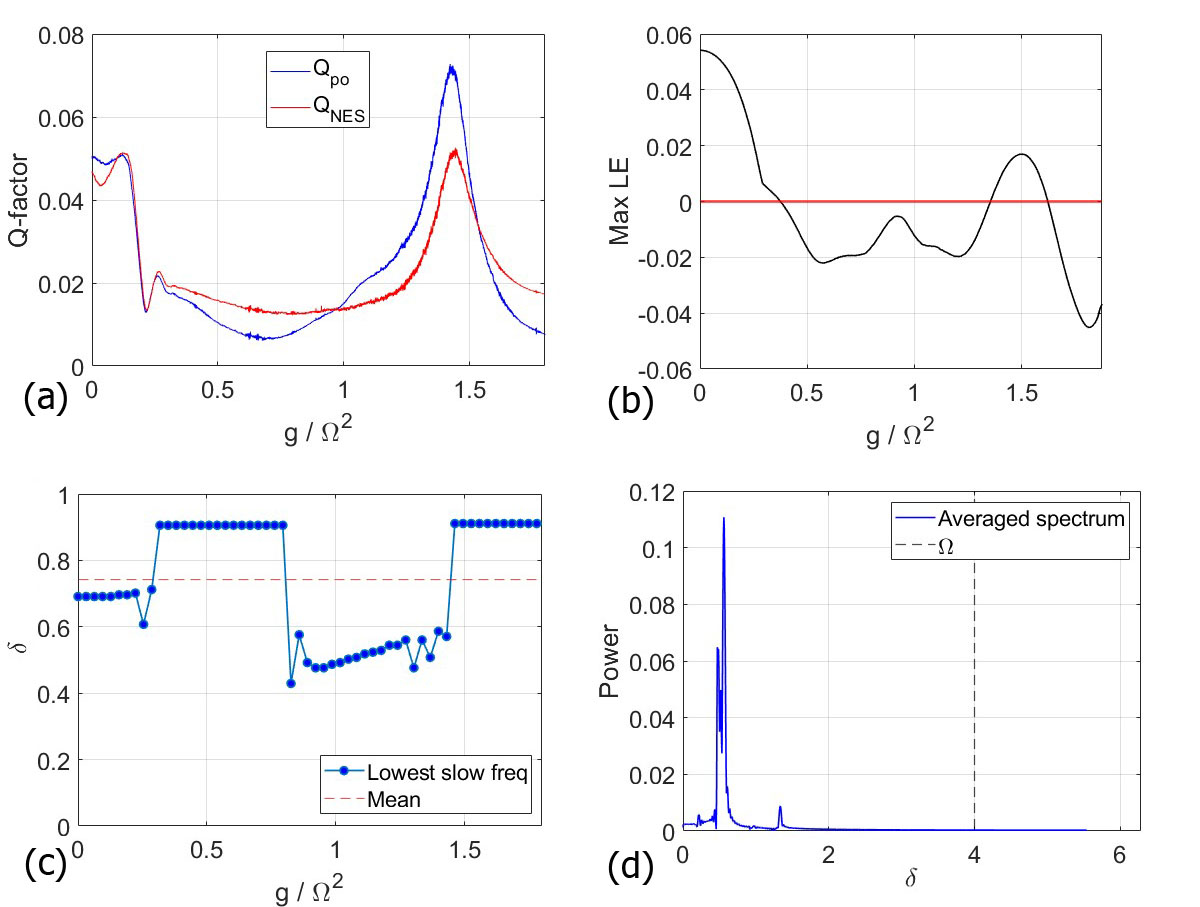}
    
\caption{\textbf{Spectral diagnostics of resonance for \(m_2=1\).} Panel (a) shows the spectral $Q$–factor  of the Van der Pol oscillator, $Q_{\mathrm{PO}}$, and that of the NES relative motion, $Q_{\mathrm{NES}}$, as functions of $g/\Omega^{2}$. Panel (b) shows the maximum Lyapunov exponent (Max LE) of the Van der Pol oscillator. Panel (c) shows the extracted lowest slow frequency, $\delta(g)$, over the $g$ sweep; the markers denote the estimates for each value of $g$, and the dashed line indicates the mean value. Panel (d) shows the averaged FFT of the demodulated \emph{weak-signal} channel (with the carrier removed) as a function of the slow frequency $\delta$, revealing a dominant slow component at $\delta\simeq 0.56~\mathrm{rad/s}$. The vertical dashed line marks the fast carrier $\Omega$ for reference. Parameters: $\alpha_{1}=0.1$, $\alpha_{2}=0.2$, $\Omega=4$, $m_{1}=1$, and $\delta=0.56~\mathrm{rad/s}$ used in the weak–signal FFT.
}
    \label{fig:2}
\end{center}
\end{figure}

\begin{figure}[htp]
\begin{center}
\includegraphics[width=15.5cm, ,clip=true]{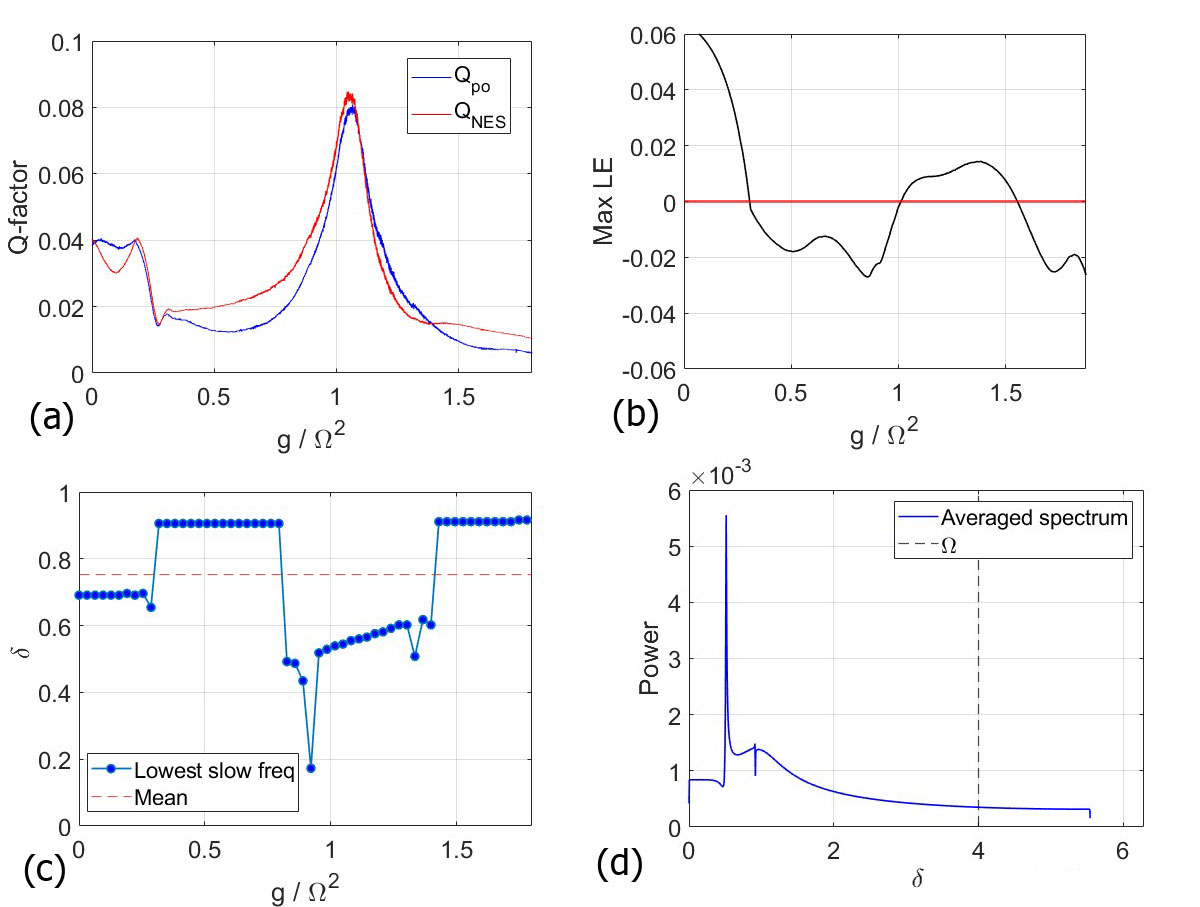}

\caption{\textbf{Spectral diagnostics of resonance for $m_2=0.8$.} Panel (a) shows the spectral $Q$-factor of the Van der Pol oscillator, $Q_{\mathrm{PO}}$, and that of the NES relative motion, $Q_{\mathrm{NES}}$, as functions of $g/\Omega^{2}$. Panel (b) shows the maximum Lyapunov exponent (Max LE) of the Van der Pol oscillator. Panel (c) shows the extracted lowest slow frequency, $\delta(g)$, over the $g$ sweep; the markers denote the estimates for each value of $g$, and the dashed line indicates the mean value. Panel (d) shows the averaged FFT of the demodulated \emph{weak-signal} channel (with the carrier removed) as a function of the slow frequency $\delta$, revealing a dominant slow component at $\delta \simeq 0.51~\mathrm{rad/s}$. The vertical dashed line marks the fast carrier frequency $\Omega$ for reference. Parameters: $\alpha_{1}=0.1$, $\alpha_{2}=0.2$, $\Omega=4$, $m_{1}=1.0$, and $\delta=0.51~\mathrm{rad/s}$ used in the weak-signal FFT.}
    \label{fig:3}
\end{center}
\end{figure}

\begin{figure}[htp]
\begin{center}
\includegraphics[width=15.5cm, ,clip=true]{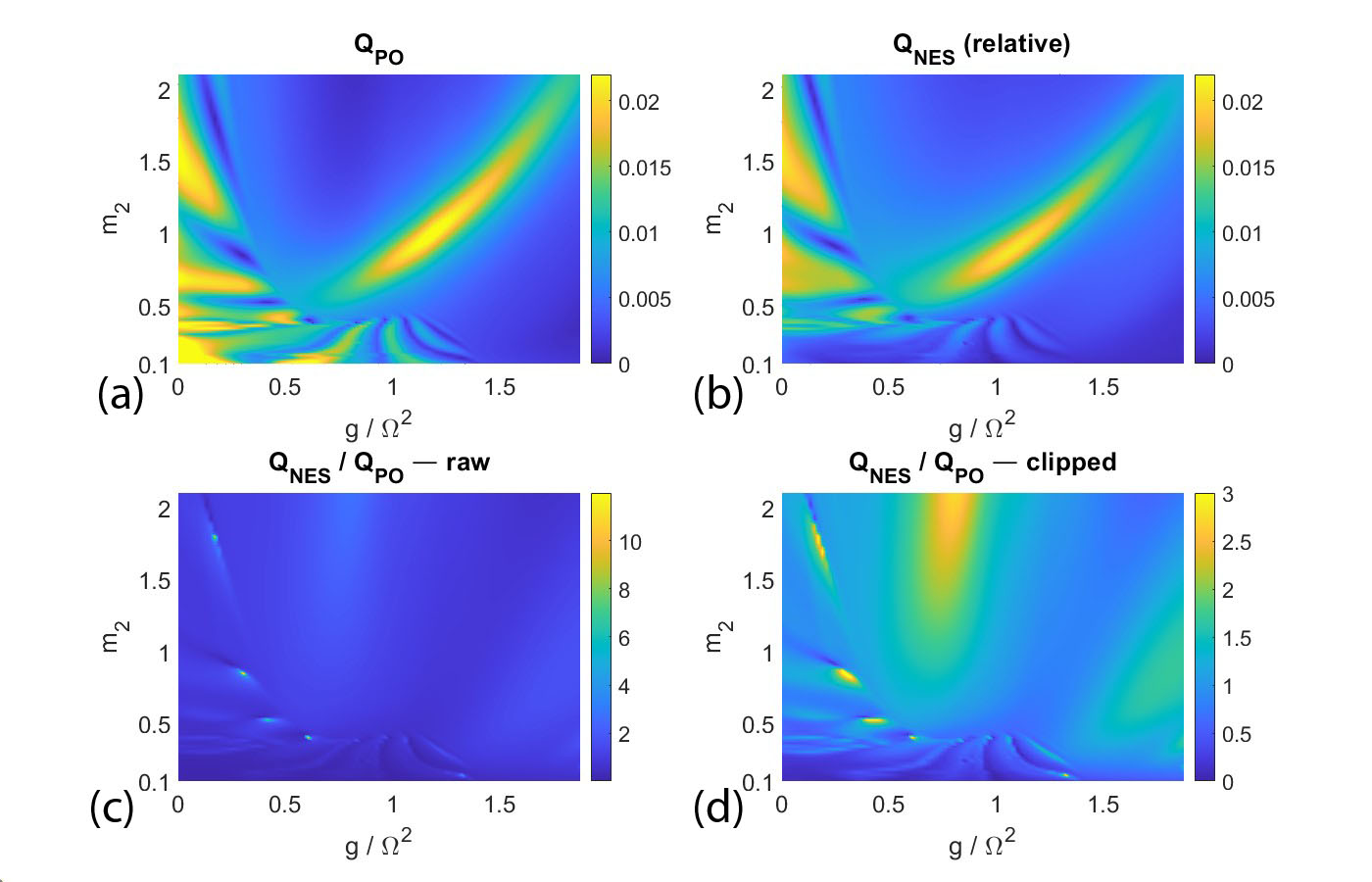}
    
\caption{\textbf{$Q$-factor maps and NES response ratio.} Panels (a) and (b) show the $Q$-factors of the Van der Pol oscillator and of the NES relative motion, respectively, in the $(m_2,\, g/\Omega^2)$ parameter plane. Panel (c) shows the ratio $Q_{\mathrm{NES}}/Q_{\mathrm{PO}}$, while panel (d) shows the same quantity clipped to the interval $[0,3]$ in order to highlight the physically relevant structure. Panels (a) and (b) share the same colormap limits. Figures~\ref{fig:2}(a) and \ref{fig:3}(a) correspond to slices of panel (a) at $m_2=1$ and $m_2=0.8$, respectively. The fixed parameters are $\alpha_1=0.1$, $\alpha_2=0.2$, $\Omega=4$, and $m_1=1$.}    
    \label{fig:4}
\end{center}
\end{figure}

\begin{figure}[htp]
\begin{center}
\includegraphics[width=15.5cm, ,clip=true]{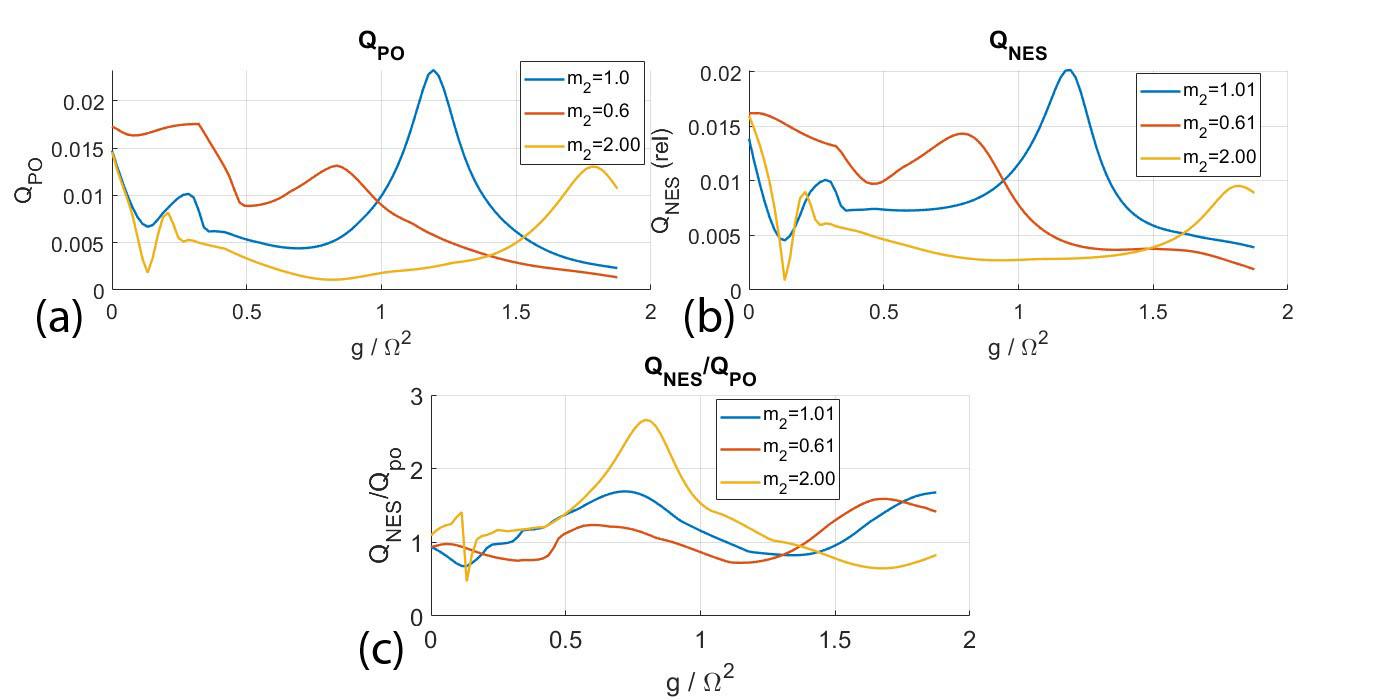}
    \caption{\textbf{$Q$-factor response for varying NES mass.} Panels (a) and (b) show the $Q$-factors of the Van der Pol oscillator and of the NES relative motion, respectively, while panel (c) shows the ratio $Q_{\mathrm{NES}}/Q_{\mathrm{PO}}$, all plotted as functions of $g/\Omega^2$ for three fixed values of the mass $m_2$. The fixed parameters are $\alpha_1=0.1$, $\alpha_2=0.2$, $\Omega=4$, and $m_1=1$. The position and amplitude of the resonance peaks vary with $m_2$, indicating how the NES mass tunes the resonance condition and affects the efficiency of energy transfer. In particular, the figure highlights how the resonance peaks shift and change amplitude across panels, reflecting the sensitivity of the response and energy transfer to the mass ratio.}
    \label{fig:5}
\end{center}
\end{figure}

\begin{figure}[htp]
\begin{center}
\includegraphics[width=15.5cm, ,clip=true]{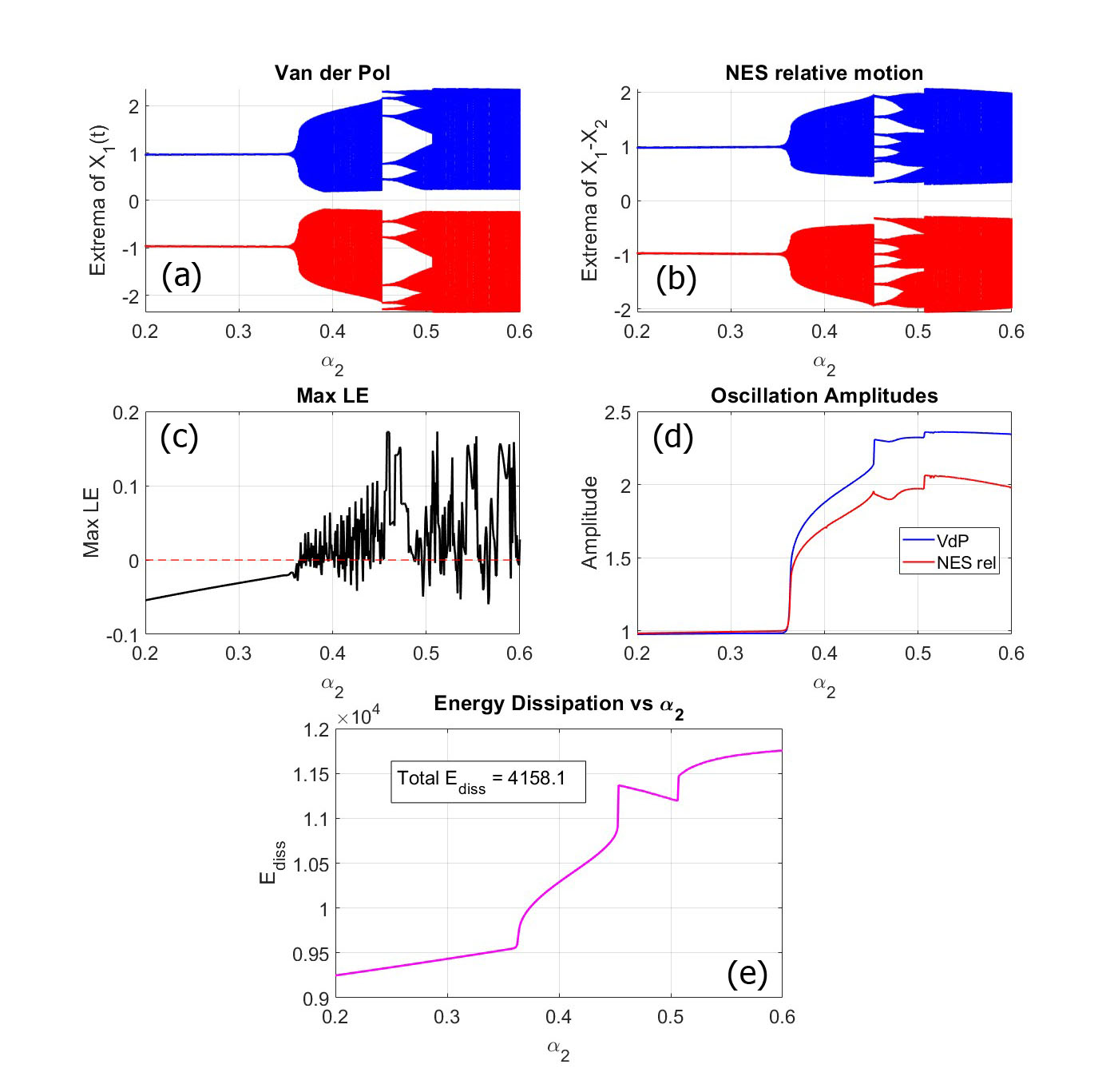}

\caption{\textbf{Extrema diagrams and energy dissipation for varying $\alpha_2$.} Panel (a) shows the maxima and minima diagram of the Van der Pol primary oscillator (PO), while panel (b) shows the corresponding extrema diagram of the NES relative motion, $x_1-x_2$. Both diagrams are constructed from the maxima and minima of the steady-state response. Panel (c) shows the maximum Lyapunov exponent (Max LE) of the system. Panel (d) shows the oscillation amplitudes of the primary oscillator and of the NES relative motion. Panel (e) reports the dissipated energy,
$E_{\mathrm{diss}} = \int_0^{t_{\max}} \gamma_2 (\dot{x}_1(t)-\dot{x}_2(t))^2 \,\mathrm{d}t$ computed step by step, together with its total value integrated over the full range of $\alpha_2$. The fixed parameters are $g=16$, $\Omega=4$, and $\alpha_1=0.1$.}
    \label{fig:6}
\end{center}
\end{figure}
\begin{figure}[htp]
\begin{center}
\includegraphics[width=15.5cm, clip=true]{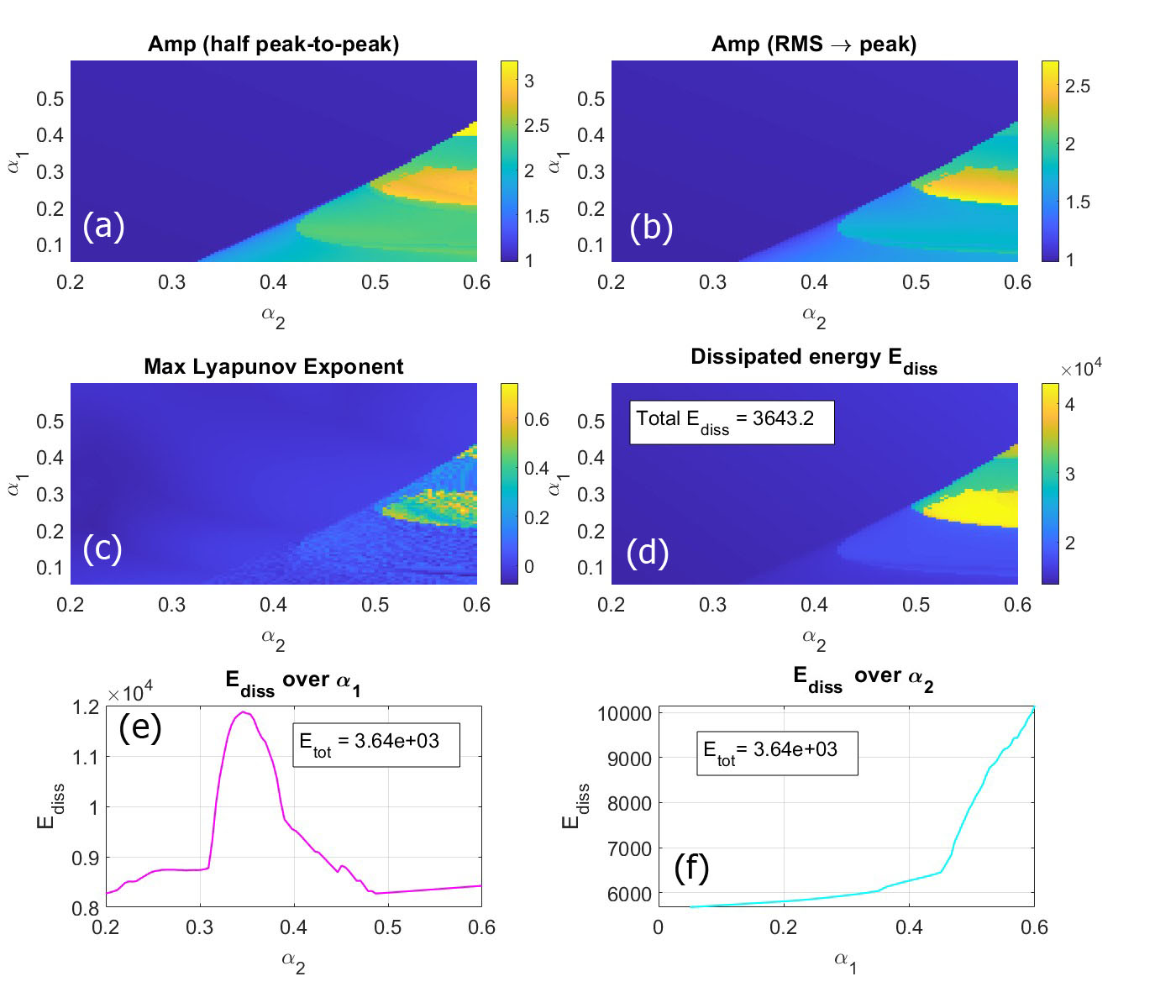}
\caption{\textbf{Amplitude, stability, and dissipation in the $(\alpha_1,\alpha_2)$ plane.} Panels (a) and (b) show the amplitudes of the Van der Pol oscillator and of the NES relative motion, respectively. Panel (c) shows the maximum Lyapunov exponent of the full coupled dynamics. Panel (d) shows the dissipated energy, defined as $E_{\mathrm{diss}} = \int_0^{t_{\max}} \gamma_2 (\dot{x}_1(t)-\dot{x}_2(t))^2 \,\mathrm{d}t$, computed for each pair $(\alpha_1,\alpha_2)$. Panels (e) and (f) show the dissipated energy integrated over $\alpha_1$ as a function of $\alpha_2$, and integrated over $\alpha_2$ as a function of $\alpha_1$, respectively. In both panels, the text box reports the corresponding total dissipated energy $E_{\mathrm{tot}}$. The parameter values are $g=15$ and $\Omega=4$.}
    \label{fig:7}
\end{center}
\end{figure}

\section{\label{sec:numerical}Vibrational Internal Resonance: Numerical analysis}

\subsection{Resonance as a function of the forcing amplitude \texorpdfstring{$g$}{g}}

This section presents the results of the numerical analysis. The system parameters are fixed at $m_2=1.0$ in Fig.~\ref{fig:2} and $m_2=0.8$ in Fig.~\ref{fig:3}. In both figures, panel (a) shows the \emph{spectral} $Q$-factor of the primary oscillator and of the NES relative motion, while panel (b) shows the maximum Lyapunov exponent (Max LE) of the primary oscillator; the remaining parameter values are given in the figure captions. In both cases, the $Q$-factor exhibits a clear resonance peak, although the value of $g/\Omega^{2}$ at which it occurs depends on the mass ratio. The $Q$-factor is used here as a spectrally focused indicator of resonance, defined as the magnitude of the Fourier projection of the response onto the slow characteristic frequency. Unlike raw amplitude-based measures, this quantity isolates the coherent oscillatory energy concentrated at a specific frequency and therefore provides a robust diagnostic of resonance capture and targeted energy transfer. Regions where $Q_{\mathrm{NES}} > Q_{\mathrm{PO}}$ indicate effective energy redirection toward the nonlinear energy sink, thereby providing a practical signature of internal vibrational resonance.

From Eq.~\eqref{eq:resonance_fast_amp}, the resonance peak is located at
$(g/\Omega^2)_{\mathrm{res}} \approx 1.82$
for the parameter values $\alpha_1=0.1$, $\alpha_2=0.2$, and $\omega_0=\sqrt{k_1/m_1}=0.707$. For $m_2=0.8$ (i.e., $\epsilon=0.8$), no simple closed-form expression is available for the resonance peak. In this case, the resonant value must be determined numerically from the explicit $1{:}1$ resonance condition $\omega_p=\omega_n$,
where $\omega_p^2$ and $\omega_n^2$ are given by Eqs.~\eqref{eq:effetive_po_freq} and~\eqref{eq:effetive_nes_freq}, respectively. This yields $(g/\Omega^2)_{\mathrm{res}} \approx 1.83$

The discrepancy between the resonance peak obtained from the numerical process and the value predicted by the analytical estimate (i.e., $\omega_p=\omega_n$) arises from the algorithm used to compute the $Q$–factor.  In standard numerical practice, the Fourier spectral projection is evaluated at the frequency of the weak signal (see e.g.~\cite{landa2000vibrational,baltanas2003experimental}), rather than at the unperturbed natural frequency $\omega_0$.  In the present problem, there is no externally imposed weak signal at a fixed frequency; instead, the NES plays the role of an \emph{internal} weak excitation to the primary oscillator (and vice versa), whose characteristic frequency (say $\delta$) itself evolves with the applied high–frequency drive.  Consequently, the weak–signal frequency $\delta$ is not stationary, but changes as the fast forcing is varied (see Figs.~\ref{fig:2}(c) and~\ref{fig:3}(c); discussed later), and a static Fourier projection at $\omega_0$ will generally miss the true resonant peak.  

To account for this effect, the weak component must be captured dynamically. In practice, we estimate the slowly varying carrier by averaging out the fast component of the frequency and taking the power spectra at different slow frequencies by FFT, which reveals the actual slow component of the frequency at a particular value of high frequency signal strength $g$. In Figs.~\ref{fig:2}(d) and~\ref{fig:3}(d), the values of $\delta$ are obtained from FFT after discarding the post-transient, low-pass filtered response, thereby isolating the slow modulation frequency after removing the fast forcing component at $\Omega$, and from which the dominating slow component frequencies are identified. For $m_2=1$, a secondary spectral peak appears close to the dominant slow frequency, indicating modal splitting due to near-symmetric coupling. In contrast, for $m_2=0.8$, this secondary peak disappears and a single dominant frequency emerges, reflecting stronger energy localization and cleaner resonance capture. The values are $\delta=0.56$ for $m_2=1$ and  $\delta=0.51$ for $m_2=0.8$ are obtained numerically from these figures. These two values of $\delta$'s are used to compute the $Q$ factors of both primary oscillator and NES, shown in respective Figs.~\ref{fig:2}(a) and~\ref{fig:3}(a). The algorithm for the determination of Q is discussed below. This dynamic tracking of the weak-signal frequency $\delta$, on which the spectral projection of the system is based, is one of the key aspects of \emph{Vibrational internal resonance}. Using the dynamically extracted slow frequency $\delta$ as an effective frequency scale, Eq.~\eqref{eq:resonance_fast_amp} suggests the modified estimate
%So, in Eq.~\eqref{eq:resonance_fast_amp} instead of static $\omega_0$, the newly predicted dynamically adjusted frequency is replaced, i.e., 
\begin{equation}
    \left(\dfrac{g}{\Omega^2}\right)_{res}\approx\delta\sqrt{\dfrac{2}{3(\alpha_2-\alpha_1)}},
    \label{eq:resonance_fast_amp_delta}
\end{equation}
that yields the resonance peak $(g/\Omega^2)_{\mathrm{res}} \approx 1.44$ for $\delta=0.56$, $m_2=1$ and $(g/\Omega^2)_{\mathrm{res}} \approx 1.31$ for  $\delta=0.51$,$m_2=0.8$. The values of the resonance peak obtained are in good agreement with Figs.~\ref{fig:2}(a) and ~\ref{fig:3}(a). The analytical prediction of $\delta$ for this framework constitutes a crucial direction for future research. 

We evaluate a spectral $Q$--factor by projecting the time series onto the carrier frequency $\delta$ and normalizing by a robust peak scale from the coupled channel. For finite records, the spectral projection is computed by a composite trapezoid (or the nearest Discrete Fourier Transform bin); see Appendix~\ref{app:qfactor}. With $T$ the total observation window,
\begin{equation}
Q_{\mathrm{PO}}
=\frac{|\hat Y(\delta)|}{\delta^2\,\max_{t}|x_1(t)-x_2(t)|+\varepsilon},
\label{eq:qpo_main}
\end{equation}
\begin{equation}
Q_{\mathrm{NES}}
=\frac{|\widehat{Y_{\mathrm{rel}}}(\delta)|}{\delta^2\,\max_{t}|x_1(t)|+\varepsilon},
\label{eq:qnes_main}
\end{equation}
where $\varepsilon=10^{-12}$ prevents division by zero. Here $\hat Y(\delta)$ and $\widehat{Y_{\mathrm{rel}}}(\delta)$ are the complex spectral projections defined in Appendix~\ref{app:qfactor}, and $\delta^2$ provides the acceleration-type scaling used to interpret the peak displacement as an effective forcing amplitude.

We use the full time series along with the transients as the resonance capture and energy transfer occur primarily during this stage. In a steady regime, the NES is entrained by the primary oscillator, and the resonance profile diminishes. Consequently, the full trajectory yields a physically meaningful $Q$-factor. Using the same procedure for both oscillators enables consistent comparison.
 
Moreover, we compute the NES $Q$–factor from its \emph{relative} motion $x_1(t)-x_2(t)$ rather than from the absolute displacement $x_2(t)$. The NES interacts with the primary oscillator solely through the relative coordinates: the coupling and damping depend on $(x_1-x_2)$ and $(\dot{x_1}-\dot{x_2})$, and the dissipated energy satisfies $E_{\mathrm{diss}}\propto \int (\dot{x_1}-\dot{x_2})^2\,\mathrm{d}t$. Using $x_2(t)$ alone would overstate the NES response by including entrained components of the primary motion, whereas $x_1-x_2$ isolates the absorber’s effective dynamics that mediate energy transfer. This choice also aligns the $Q$–factor with our energy metric and yields a physically interpretable, like–for–like comparison with the primary oscillator response.

The maximum Lyapunov exponent (Max LE) does not exhibit any systematic correspondence with the resonance peaks in Figs.~\ref{fig:2}(b) and \ref{fig:3}(b). This indicates that the onset of resonance is not tied to the emergence of chaos: resonance capture may occur in both regular and weakly chaotic regimes. In contrast to the $Q$-factor analysis, which is based on the full trajectory, the Lyapunov analysis characterizes the long-term attractor and is therefore performed only on the \emph{steady-state} dynamics, after discarding an initial transient. Unless stated otherwise, we remove the first $t_{\mathrm{tr}}=0.4\,T$ of each time series, where $T$ denotes the total simulation time, and compute the Lyapunov exponents over the remaining interval $T_{\mathrm{LE}}=T-t_{\mathrm{tr}}$. The Lyapunov exponents are computed on the full four-dimensional state space using the Benettin--QR algorithm with a numerically evaluated Jacobian.

Figures~\ref{fig:2}(c) and \ref{fig:3}(c) show the slow-frequency plateaus as the fast-forcing strength, $g/\Omega^2$, is varied. The dynamically extracted slow frequency exhibits piecewise constant plateaus separated by abrupt jumps. These discontinuities correspond to switching between distinct slow-flow attractors of the nonlinear averaged system. In particular, one plateau appears near $\delta \approx 0.56$ in Fig.~\ref{fig:2}(c), while another appears near $\delta \approx 0.51$ in Fig.~\ref{fig:3}(c), precisely in the parameter regions where the $Q$-factor attains its maximum in the corresponding $Q$-factor plots shown in Figs.~\ref{fig:2}(a) and \ref{fig:3}(a). This confirms that the resonance peak is associated with the nonlinear matching condition $\omega_p \approx \omega_n$, which enables phase locking and efficient energy transfer. The observed jumps therefore mark entry into, and exit from, the resonance-capture manifold as the system undergoes bifurcation-induced transitions between different solution branches.

Figure~\ref{fig:4} maps $Q_{\mathrm{PO}}$ and $Q_{\mathrm{NES}}$ in the parameter plane $(g/\Omega^{2},m_{2})$, revealing a high-$Q$ region for both oscillators as a continuous band in the plane. This high-$Q$ region corresponds to the condition $\omega_p \approx \omega_n$, identifying the resonance-capture region with coherent NES amplification, which is consistent with the internal-resonance condition predicted by Eq.~\eqref{eq:resonance_fast_amp}. Along this curve, the $1{:}1$ internal resonance occurs due to the coincidence of effective frequencies, which enables efficient energy transfer from the primary oscillator to the NES. Panels (a,b) have the same colormap limits, while panels (c,d) show the ratio  $R=Q_{\mathrm{NES}}/Q_{\mathrm{PO}}$, where $R\gg1$ near the peak indicates dominant NES response and effective energy transfer due to suppressed primary motion. To visualize more clearly, panel (d) shows clipped data of Q ($[0,3]$), indicating the regimes $R>1$, where the NES response is dominant, $R<1$ indicates primary-oscillator-dominated response, and $R\approx1$ indicates comparable response. Further, regions with $R\approx0$ indicate the absence of phase locking and resonance capture, while the slices at $m_2=1$ and $m_2=0.8$ correspond to Figs.~\ref{fig:2} and~\ref{fig:3}, respectively.

Figure~\ref{fig:4} shows that the NES mass $m_2$ controls resonance capture through the parameter $\epsilon$, shifting the effective NES frequency, and modifying the condition $\omega_p=\omega_n$. This governs the energy transfer efficiency and system stability. Over $m_2\in[0.1,2]m_1$, the response structure of $Q_{\mathrm{PO}}$ and $Q_{\mathrm{NES}}$ varies with $(g/\Omega^2,m_2)$. The response is weak for $m_2\lesssim0.6m_1$ and becomes stronger and more continuous near $m_2\sim m_1$, with extended regions of $R>1$. For $m_2\gtrsim1.5m_1$, resonance peaks shift to lower $g/\Omega^2$ and shrink. The high-$Q$ regime collapses under weak forcing, showing that capture is robust near $m_2\sim m_1$ and declines for very light or heavy NES due to inertia mismatch. Figure~\ref{fig:5} further illustrates the effect of $m_2$ on resonance capture through $Q$-factors of the primary, NES, and their ratio.

\subsection{Resonance as a function of the parameters \texorpdfstring{$\alpha_1,\alpha_2$}{alpha1, alpha2}}

We now examine the $(\alpha_1,\alpha_2)$ plane, which governs detuning, backbone curves, and energy pumping with fixed forcing and NES masses $m_2$. We then quantify how $(\alpha_1,\alpha_2)$ affect response regimes and pumping efficiency, keeping $(k_1,\gamma_1,\gamma_2)$ fixed. This isolates the role of nonlinear stiffness, since $k_1$ rescales frequency and $\gamma_{1,2}$ determine dissipation and dissipative coupling.

Figure~\ref{fig:6} shows bifurcation diagrams, amplitudes, Max LE, and dissipation vs.\ $\alpha_2$, revealing a transition for $0.3\lesssim\alpha_2\lesssim0.4$ where the dynamical regime changes. The Max LE is computed for the full coupled dynamics, namely $(x_1,\dot x_1,x_2,\dot x_2)$, using the steady-state portion of the trajectories after discarding the transient.

For both the Max-LE estimate and the bifurcation diagrams, we analyze the asymptotic response: the transient is discarded, and only the steady state is sampled. In parallel with the amplitude curves and the Max-LE curve shown in Fig.~\ref{fig:6}, we quantify the net energy absorbed by the NES through viscous dissipation in the coupling damper, namely, the element proportional to $(\dot x_1-\dot x_2)$. This quantity is computed over the \emph{entire} simulated time series (transient plus steady state):
\begin{equation}
  E_{\mathrm{diss}} \;=\; \int_{0}^{t_{\max}} \gamma_{2}\,\bigl(\dot x_1(t)-\dot x_2(t)\bigr)^{2}\,\mathrm{d}t.
  \label{eq:Ediss}
\end{equation}
Since it depends only on the relative velocity, it measures the energy dissipated by the coupling damper, that is, the energy irreversibly removed from the coupled primary-oscillator--NES system through the relative-motion dissipation channel.

As argued in the discussion of the $Q$-factor, resonance capture, and the associated irreversible energy transfer, occurs predominantly during the transient stage. Restricting the analysis to the steady state would therefore substantially underestimate the absorbed energy, since the NES eventually locks to the primary oscillator and the additional dissipation becomes negligible. The use of $x_1(t)-x_2(t)$, rather than $x_2(t)$ alone, is essential: the NES interacts with the primary oscillator through the relative spring and damper, so Eq.~\eqref{eq:Ediss} directly quantifies the extent to which the absorber is engaged. In panel (e), the values of $\alpha_2$ that yield large $E_{\mathrm{diss}}$ coincide with those producing large-amplitude oscillations in panels~(a) and~(b), as well as with the dynamical reorganization indicated by the Max LE in panel~(c).

We next extend the analysis to the two-parameter maps shown in Fig.~\ref{fig:7}. Panels~(a) and~(b) report the steady-state amplitudes, panel~(c) shows the Max LE computed from the steady-state dynamics, and panel~(d) reports the dissipated energy, $E_{\mathrm{diss}}$,
computed over the \emph{full} time series in order to include the transient energy-pumping stage. Panels~(e) and~(f) display one-dimensional projections of $E_{\mathrm{diss}}(\alpha_1,\alpha_2)$ obtained by integrating over the complementary parameter: panel~(e) shows $E^{(\alpha_2)}(\alpha_1)$, integrated over $\alpha_2$, whereas panel~(f) shows $E^{(\alpha_1)}(\alpha_2)$, integrated over $\alpha_1$.

These marginals are computed as
\begin{equation}
\label{eq:Marginal}
\begin{aligned}
\textbf{Integrated over }\alpha_1:\quad
E^{(\alpha_1)}(\alpha_2)
&=\int_{\alpha_{1,\min}}^{\alpha_{1,\max}} E_{\mathrm{diss}}(\alpha_1,\alpha_2)\,\mathrm{d}\alpha_1,\\
\textbf{Integrated over }\alpha_2:\quad
E^{(\alpha_2)}(\alpha_1)
&=\int_{\alpha_{2,\min}}^{\alpha_{2,\max}} E_{\mathrm{diss}}(\alpha_1,\alpha_2)\,\mathrm{d}\alpha_2.
\end{aligned}
\end{equation}

Their trends mirror the high-amplitude regions in the maps, and the text-box values of $E_{\mathrm{tot}}$ agree across the different constructions, confirming the internal consistency of the amplitude, Lyapunov exponents, and energy-based indicators. The total energy is computed as
\begin{equation}
E_{\mathrm{tot}}
=\int_{\alpha_{2,\min}}^{\alpha_{2,\max}}
  \int_{\alpha_{1,\min}}^{\alpha_{1,\max}}
  E_{\mathrm{diss}}(\alpha_1,\alpha_2)\,\mathrm{d}\alpha_1\,\mathrm{d}\alpha_2.
\end{equation}

Figure~\ref{fig:7} shows that $(\alpha_1,\alpha_2)$ control NES performance, with significant oscillations confined to a wedge-shaped corridor in parameter space. For $\alpha_2>\alpha_1$, the primary response increases sharply, while the Max LE map shows mostly weakly chaotic or near-regular dynamics with a narrow chaotic region near the upper boundary. Thus, the strong response is primarily associated with nonlinear tuning between $\alpha_1$ and $\alpha_2$, rather than being driven by the onset of strongly chaotic dynamics, with the dissipated energy map confirming maximal energy removal in the same corridor.

Panels (e,f) show marginals of $E_{\mathrm{diss}}$ over $\alpha_1$ and $\alpha_2$ (Eq.~\eqref{eq:Marginal}), with panel (e) indicating peak dissipation at intermediate $\alpha_2$ near optimal resonance. Panel (f) shows that larger $\alpha_1$ enhances energy removal, consistent with stronger pumping, and $E_{\mathrm{tot}}=\iint E_{\mathrm{diss}}\,\mathrm{d}\alpha_1\,\mathrm{d}\alpha_2$ matches the marginals, confirming consistency of the maps.

\begin{figure}[htp]
\begin{center}
\includegraphics[width=15.5cm, ,clip=true]{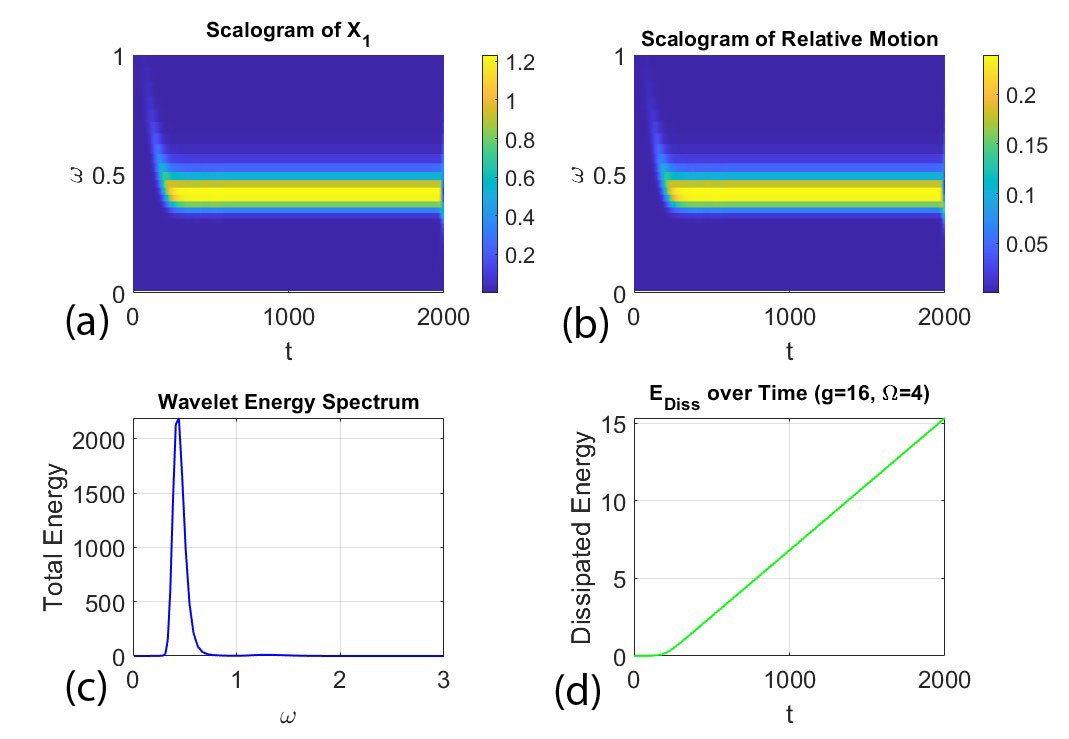}
       \caption{\textbf{Wavelet analysis and energy dissipation during resonance capture.} Panel (a) shows the scalogram of the primary displacement $x_1(t)$, computed using the continuous wavelet transform (Morlet/``amor''). The colormap represents $|\mathrm{CWT}|^{2}$, and the vertical axis corresponds to the angular frequency $\omega$ (rad/s). After a short transient, the energy concentrates in a narrow frequency band, indicating carrier capture. Panel (b) shows the scalogram of the NES relative motion, $x_1(t)-x_2(t)$; the same time-frequency ridge is observed, confirming synchronized capture in the interaction channel. Panel (c) shows the time-integrated wavelet energy spectrum, $\int |\mathrm{CWT}(\omega,t)|^{2}\,\mathrm{d}t$, for $x_1(t)$, revealing a sharp peak at the captured frequency $\omega$. Panel (d) shows the cumulative dissipated energy,
$E_{\mathrm{diss}}(t)=\int_{0}^{t}\gamma_{2}\,[\dot x_1(\tau)-\dot x_2(\tau)]^{2}\,\mathrm{d}\tau,$
which increases nearly linearly after the transient as energy is pumped into the relative damper. The parameters in the effective equations are $k_{1}=0.5$, $\gamma_{1}=\gamma_{2}=0.3$, $\alpha_{1}=0.1$, $\alpha_{2}=0.38$, and $m_{1}=1$.}
    \label{fig:8}
\end{center}
\end{figure}

\begin{figure}[htp]
\begin{center}
\includegraphics[width=15.0cm, ,clip=true]{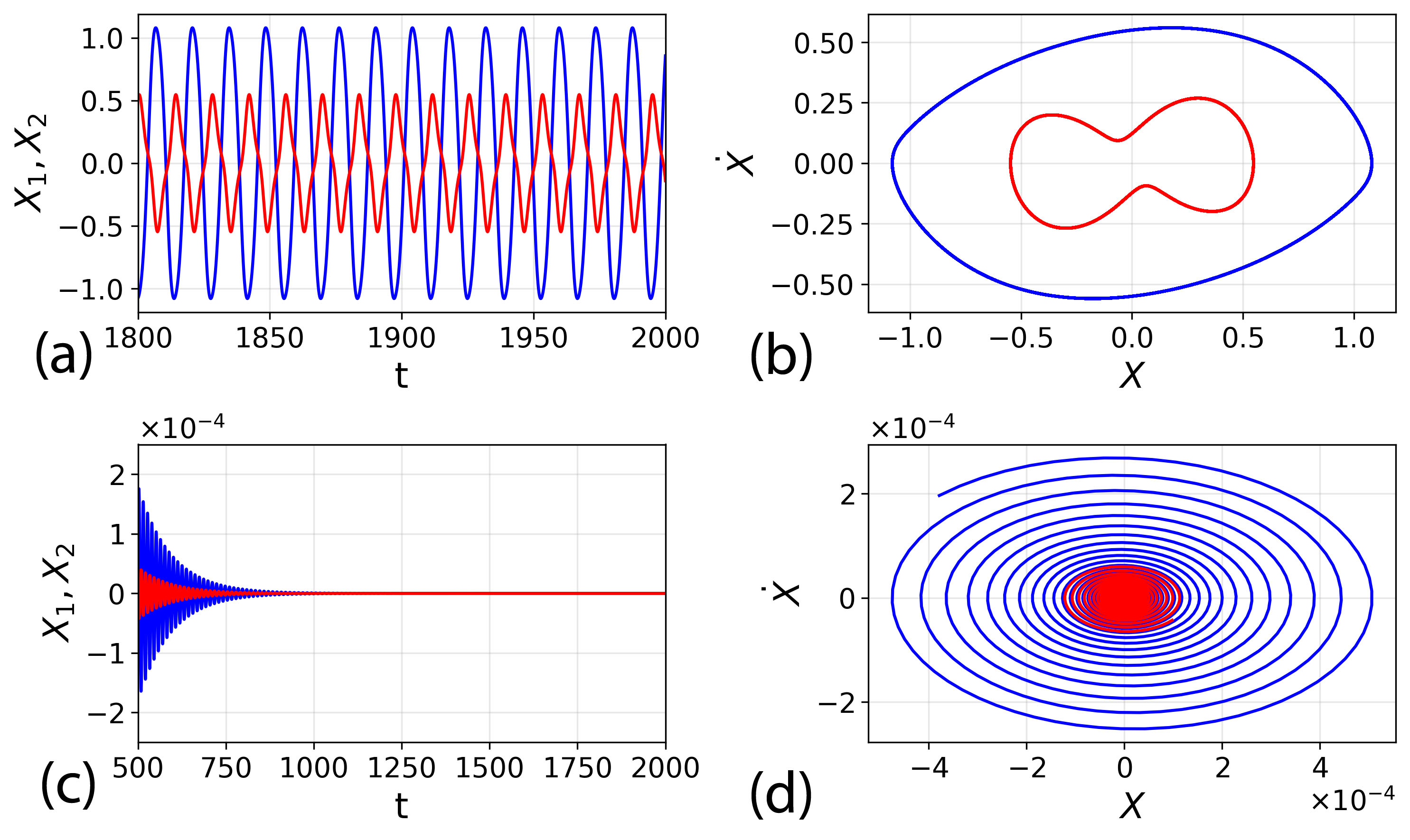}

\caption{\textbf{Control of oscillatory instability through a Hopf bifurcation.} The figure illustrates the Hopf bifurcation in the effective averaged dynamics [Eqs.~\eqref{eq:effective_dyn_PO}--\eqref{eq:effective_dyn_NES}] for $\Omega=4$. Panels (a) and (b) correspond to $g=16<g_{\mathrm{Hopf}}\approx 23$, for which $K>0$ and the system exhibits sustained limit-cycle oscillations in both the primary oscillator (blue) and the NES (red). Panel (a) shows the post-transient time evolution, while panel (b) shows the corresponding phase portraits, confirming stable limit-cycle motion. Panels (c) and (d) correspond to $g=24>g_{\mathrm{Hopf}}\approx 23$, for which $K<0$ and the system undergoes a Hopf bifurcation leading to suppression of the oscillations. The trajectories collapse onto a small-amplitude attractor, demonstrating high-frequency-forcing-induced stabilization of the effective dynamics.}
    \label{fig:9}
\end{center}
\end{figure}

\begin{figure}[htp]
\begin{center}
\includegraphics[width=17.5cm, ,clip=true]{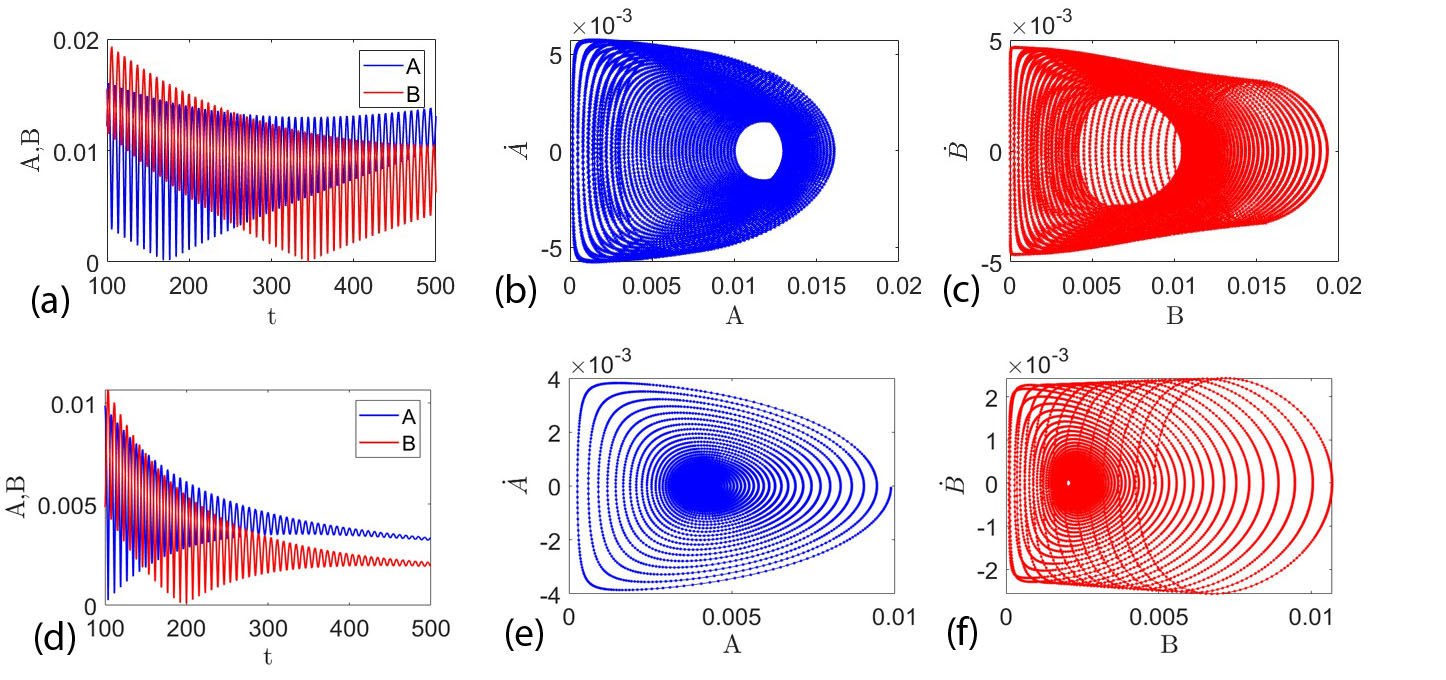}

\caption{\textbf{Suppression of oscillatory instability in the slow-flow dynamics.} The control of oscillatory instability by high-frequency forcing is further confirmed from the slow-flow amplitude dynamics governed by Eqs.~\eqref{eq:A_dynamics}--\eqref{eq:theta2_dynamics} for $\Omega=4$. Panels (a)--(c) correspond to $g=16<g_{\mathrm{Hopf}}\approx 23$, for which $K>0$ and sustained limit-cycle oscillations persist in both the primary-oscillator amplitude $A$ (blue) and the NES amplitude $B$ (red). Panel (a) shows the post-transient time evolution of $(A,B)$, while panels (b) and (c) show the corresponding phase portraits, $(A,\dot{A})$ and $(B,\dot{B})$, confirming stable periodic motion. Panels (d)--(f) correspond to $g=24>g_{\mathrm{Hopf}}\approx 23$, for which $K<0$ and the system undergoes a Hopf bifurcation leading to suppression of the oscillations. The trajectories collapse onto a small-amplitude attractor, demonstrating stabilization of the effective dynamics induced by the high-frequency drive.}
\label{fig:10}
\end{center}
\end{figure}

\subsection{Wavelet scalogram analysis}

To analyze the signals simultaneously in time and frequency, we use the continuous wavelet transform, which plays a role analogous to that of a sliding-window Fourier transform. This allows us to track the evolution of the vibration frequency during the onset of energy capture by the NES. The resulting representation, namely the scalogram, shows which frequencies are present at each instant, while the corresponding integrated wavelet spectrum quantifies how much energy is distributed across frequencies. Figure~\ref{fig:8} shows the results for the effective model with $g=16$ and $\Omega=4$.

Panels~\ref{fig:8}(a) and~\ref{fig:8}(b) show that both the primary displacement $x_1(t)$ and the relative NES motion $x_1(t)-x_2(t)$ rapidly concentrate onto a narrow frequency band after a short transient. This dominant band is consistent with the analytically predicted effective frequency of the primary oscillator (see Sec.~A), supporting the interpretation that the high-frequency drive tunes the system into a $1{:}1$ internal-resonance condition. The corresponding energy spectra, $E_{x_1}(\omega)$ and $E_{x_1-x_2}(\omega)$, shown in panel~\ref{fig:8}(c), exhibit sharp peaks, indicating a spectrally focused long-time response. Finally, the cumulative dissipated energy, $E_{\mathrm{diss}}(t)$, shown in panel~\ref{fig:8}(d), increases steadily once the frequency band is established, consistent with sustained energy pumping through the relative damper.

\subsection{Instability control through Hopf bifurcation}
Apart from the TET control through NES, the high-frequency forcing can be used to mitigate the instability of the primary oscillator by regulating the limit-cycle oscillation via \textit{Hopf bifurcation}. The characteristics of the oscillation have been verified numerically by the effective dynamics Eqs.~\eqref{eq:effective_dyn_PO} and \eqref{eq:effective_dyn_NES} for different values of $g$. For reference, Fig.~\ref{fig:9} compares the steady--state time histories and phase portraits of the primary oscillator and the NES (\(X_2\)) at \(\Omega=4\), both on and beyond the \(Q\)--ridge. According to Eq.~\eqref{eq:hopf_point}, the Hopf transition occurs at \(g_{\mathrm{hopf}}\approx 23\). For \(g=16<g_{\mathrm{hopf}}\), the primary oscillator evolves on a pronounced limit cycle and the NES response is also sustained, yielding a phase--locked regime consistent with resonance capture (Fig.~\ref{fig:9} (a),(b)). In contrast, when the high-frequency drive strength is increased to \(g=24>g_{\mathrm{hopf}}\), the oscillations of the primary oscillator are markedly reduced and the NES response collapses as well, with the trajectories converging to a small--amplitude attractor (Fig.~\ref{fig:9} (c),(d)). These results indicate that the high frequency drive can be used as an effective stability--control mechanism: beyond the ridge, the loss of capture suppresses the large--amplitude response in both coordinates. To validate the full-system results mechanistically, we derive the modulation (slow--flow)
equations for the envelopes and phases (Eqs.~\eqref{eq:A_dynamics}--\eqref{eq:theta2_dynamics})
using the ansatz
\begin{equation}
\label{eq:slowflow_ansatz}
\begin{aligned}
X_1(t) &\approx A(T)\,\cos\!\big(\omega_p t+\theta_{1}(T)\big),\\
X_2(t) &\approx B(T)\,\cos\!\big(\omega_p t+\theta_{2}(T)\big),
\end{aligned}
\end{equation}
where $A(T)$ and $B(T)$ are the modulation envelopes (modal amplitudes) and $\theta_{1,2}(T)$ are the slow phases of the primary and relative (NES) motions, respectively. Here $T=\varepsilon \tau$ denotes
the slow time, and $\omega_p$ is the high-frequency--tuned frequency derived in Sec.~\ref{model}.
Figure~\ref{fig:10} reports representative steady-state slow--flow responses selected on the high-$Q$
ridge ($g=16$) and beyond it ($g=24$). For $g=16$ (top row), the envelopes $A(t)$ and $B(t)$ exhibit
net growth before saturating to sustained modulation oscillations (Fig.~\ref{fig:10} (a)), and the
corresponding portraits $(A,\dot A)$ and $(B,\dot B)$ converge to a closed orbit
(Fig.~\ref{fig:10} (b),(c)), i.e., a finite-amplitude limit cycle of the averaged system consistent with
resonance capture. In contrast, for $g=24$ (bottom row), the envelopes decay toward small steady values
(Fig.~\ref{fig:10} (d)) and the trajectories in $(A,\dot A)$ and $(B,\dot B)$ spiral to a stable fixed
point (Fig.~\ref{fig:10} (e),(f)), indicating loss of capture outside the ridge. Thus, the slow-flow interprets the $Q$-maps dynamically: within the capture corridor, a finite-amplitude attractor exists, while outside the system relaxes to equilibrium; Fig.~\ref{fig:10} shows asymptotic behavior with parameters chosen $\mathcal{O}(\varepsilon^{2})$, consistent with $\mathcal{O}(\varepsilon)$ accuracy.

\section{Conclusions}

We investigate vibrational internal resonance in a Van der Pol oscillator coupled to a nonlinear energy sink (NES), where slow-fast averaging shows that tuning $g/\Omega^2$ enables resonance capture, which is characterized by a spectrally focused Q-factor based on the FFT of the slow-envelope frequency $\delta$. 
In $(g/\Omega^{2},m_{2})$ and $(\alpha_{1},\alpha_{2})$ spaces, resonance capture is robust with regions where $Q_{\mathrm{NES}}>Q_{\mathrm{PO}}$, and targeted energy transfer is most effective for $0.6\!\lesssim\!\mu\!\lesssim\!2$, degrading for very light or heavy NES. For fixed forcing, the $(\alpha_{1},\alpha_{2})$ plane exhibits a band of amplitude transitions that is not mirrored in the Lyapunov exponents, which suggests that \emph{transient capture} is mostly separated from the \emph{asymptotic} dynamics.

The wavelet transformation (Fig.~\ref{fig:8}) shows a dominant slow band (from the effective dynamics) near $\omega_p\approx\omega_n\sim0.5$ during transients, which is consistent with $Q$-factor peaks and growth of $E_{\mathrm{diss}}(t)$. Overall, the study provides a physically grounded mechanism for active high-frequency tuning of NES performance, with potential relevance to vibration mitigation and energy-harvesting applications based on internal resonance. Finally, we have demonstrated the instability control of the effective system via Hopf bifurcation through the parameter $K=1-g^{2}/(2\Omega^{4})$, where tuning $g/\Omega^{2}$ switches the system from limit-cycle oscillations ($K>0$) to oscillation suppression ($K<0$), providing a mechanism for stability control.

Robust TET occurs for $0.6\lesssim\mu\lesssim2$ with the strongest capture near $m_{2}\sim m_{1}$, while $\alpha_{2}\in[0.30,0.40]$ maximizes the $Q$-ridge and dissipation. In contrast, $\alpha_{1}$ appears to play a secondary role. Future work should extend the analysis to multi-DOF systems and experimental validation.

\section{Acknowledgment}
 SR is thankful to Dr. Rahul Das for the constructive discussion during the work. MC and MAFS acknowledge the support by the Spanish State Research Agency (AEI) and the European Regional Development Fund (ERDF, EU) under Project No.~PID2023-148160NB-I00 (MCIN/AEI/10.13039/ 501100011033). They are also grateful to Rey Juan Carlos University (Spain) through its Own Research Promotion and Development Program under the project NABEH, funded by the Impulso Project Funding Scheme (Project No. 2025/00014/047). SG is thankful for the financial support received from the Ministry of Education, Govt of India, towards IoE Projects Phase II for the project titled Complex Systems \& Dynamics.

\appendix

% \section{Derivation of effective dynamics: Eqns.~\eqref{eq:effective_dyn_PO} and \eqref{eq:effective_dyn_NES}.}
\section[Derivation of effective dynamics]%
{Derivation of effective dynamics: Eqs.~\texorpdfstring{\eqref{eq:effective_dyn_PO}}{(PO)} 
and~\texorpdfstring{\eqref{eq:effective_dyn_NES}}{(NES)}}
\label{appendix:A}

In this appendix, the effective slow dynamics of the system is derived. Using the method of direct partition of motion from vibrational mechanics, the original coordinates are decomposed into slow and fast components. Averaging over the fast oscillations generated by the high–frequency forcing yields the effective equations governing the slow variables $X_1$ and $X_2$. To derive Eqs.~\eqref{eq:effective_dyn_PO} and \eqref{eq:effective_dyn_NES}, at first Eqs. \eqref{eq:slow_fast} are substituted into Eqs. \eqref{eq:relative_PO} and \eqref{eq:relative_NES} that gives

\begin{equation}
    \begin{split}
        \ddot{X}_1+\ddot{\psi}_1&+\gamma_1\left(X_1^2+\psi_1^2-1\right)\left(\dot{X}_1+\dot{\psi}_1\right)+\omega_0^2\left(X_1+\psi_1\right)+\alpha_1\left(X_1^3+3X_1^2\psi_1+3X_1\psi_1^2+\psi_1^3\right)\\
&+\gamma_2\left(\dot{X}_2+\dot{\psi}_2\right)+\alpha_2\left(X_2^3+3X_2^2\psi_2+3X_2\psi_2^2+\psi_2^3\right)=g\cos\Omega t
\label{eq:A1}
    \end{split}
\end{equation}
and
\begin{equation}
    \begin{split}
        \epsilon\left(\ddot{X}_2+\ddot{\psi}_2\right)&+\gamma_2\left(\dot{X}_2+\dot{\psi}_2\right)+\alpha_2\left(X_2^3+3X_2^2\psi_2+3X_2\psi_2^2+\psi_2^3\right)=0
        \label{eq:A2}
    \end{split}
\end{equation}

By averaging Eqs.~\eqref{eq:A1} and \eqref{eq:A2} over the fast time scale (\(\Omega t\)), while the slow components $X_j$ hardly change over this time scale, the effective slow motions are obtained as

\begin{equation}
    \begin{split}
        \ddot{X}_1+\langle\ddot{\psi}_1\rangle&+\gamma_1\left(X_1^2+\langle\psi_1^2\rangle-1\right)\dot{X}_1+\gamma_1\left(X_1^2\langle\dot{\psi}_1\rangle+\langle\psi_1^2\dot{\psi}_1\rangle-\langle\dot{\psi}_1\rangle  \right)+\omega_0^2\left(X_1+\langle\psi_1\rangle\right)\\
        &+\alpha_1\left(X_1^3+3X_1^2\langle\psi_1\rangle+3X_1\langle\psi_1^2\rangle+\langle\psi_1^3\rangle\right)\\
        &+\gamma_2\left(\dot{X}_2+\langle\dot{\psi}_2\rangle\right)+\alpha_2\left(X_2^3+3X_2^2\langle\psi_2\rangle+3X_2\langle\psi_2^2\rangle+\langle\psi_2^3\rangle\right)=\langle g\cos\Omega t\rangle
\label{eq:A3}
    \end{split}
\end{equation}
and
\begin{equation}
    \begin{split}
        \epsilon\left(\ddot{X}_2+\langle\ddot{\psi}_2\rangle\right)&+\gamma_2\left(\dot{X}_2+\langle\dot{\psi}_2\rangle\right)+\alpha_2\left(X_2^3+3X_2^2\langle\psi_2\rangle+3X_2\langle\psi_2^2\rangle+\langle\psi_2^3\rangle\right)=\epsilon\left(\ddot{X}_1+\langle\ddot{\psi}_1\rangle\right)
        \label{eq:A4}
    \end{split}
\end{equation}
where the average over the fast scale is defined as $\langle\cdot\rangle=\dfrac{1}{T}\int_0^T\cdot~~\mathrm{d}(\Omega t)$. It is also assumed that $\psi$ being a harmonic in nature, $\langle\psi\rangle=\langle\dot{\psi}\rangle=\langle\ddot{\psi}\rangle=\cdots=0$ over a complete cycle. Then subtracting Eq.\eqref{eq:A3} from Eq.\eqref{eq:A1} and Eq.\eqref{eq:A4} from Eq.\eqref{eq:A2}, the corresponding fast dynamics can be expressed as

\begin{equation}
    \begin{split}
        \ddot{\psi}_1&+\gamma_1\left(X_1^2\dot{\psi}_1+\left(\psi_1^2-\langle\psi_1^2\rangle\right)\dot{X}_1+\psi_1^2\dot{\psi}_1-\langle\psi_1^2\dot{\psi}_1\rangle-\dot{\psi}_1\right)+\omega_0^2\psi_1\\
        &+\alpha_1\left(3X_1^2\psi_1+3X_1\left(\psi_1^2-\langle\psi_1^2\rangle\right)+\psi_1^3-\langle\psi_1^3\rangle\right)+\gamma_2\dot{\psi}_2\\
        &+\alpha_2\left(3X_2^2\psi_2+3X_2\left(\psi_2^2-\langle\psi_2^2\rangle\right)+\psi_2^3-\langle\psi_2^3\rangle\right)=g\cos\Omega t
        \label{eq:A5}
    \end{split}
\end{equation}
and
\begin{equation}
    \begin{split}
\epsilon\ddot{\psi}_2+\gamma_2\dot{\psi}_2+\alpha_2\left(3X_2^2\psi_2+3X_2\left(\psi_2^2-\langle\psi_2^2\rangle\right)+\psi_2^3-\langle\psi_2^3\rangle\right)=\epsilon\ddot{\psi}_1.
        \end{split}
        \label{eq:A6}
\end{equation}

According to the inertial approximation \cite{blekhman2000vibrational}, $\ddot{\psi}_j\gg\dot{\psi}_j\gg\psi_j,\psi_j^2,\psi_j^3...$, ${j=1,2}$. Thus, Eqs.~\eqref{eq:A5} and \eqref{eq:A6} can be approximated as

\begin{equation}
    \ddot{\psi}_1\approx g\cos(\Omega t)
    \label{eq:A7}
\end{equation}
and
\begin{equation}
    \ddot{\psi}_2+\dfrac{\gamma_2}{\epsilon}\dot{\psi_2}\approx g\cos(\Omega t).
    \label{eq:A8}
\end{equation}

The reason for keeping the $\dot{\psi}_2$ term in Eq.~\eqref{eq:A8} is that its contribution is significant as the coefficient of $\dot{\psi}_1$ contains $\epsilon$ in the denominator.The solutions of $\psi_1$ and $\psi_2$ is given by

\begin{equation}
    \psi_1=-\dfrac{g}{\Omega^2}\cos(\Omega t)
\end{equation}
and
\begin{equation}
    \psi_2= \frac{\epsilon g}{\Omega\sqrt{\epsilon^2\Omega^2+\gamma_2^2}}
\cos\!\left(\Omega t - \tan^{-1}\!\frac{\gamma_2}{\epsilon\Omega}\right).
\end{equation}

This leads to
\begin{equation}
    \langle\psi_1^2\rangle=\dfrac{g^2}{2\Omega^4};~~\langle\psi_2^2\rangle=\dfrac{\epsilon^2g^2}{2\Omega^2(\epsilon^2\Omega^2+\gamma_2^2)}~~\text{and}~~\langle\psi_j\rangle=\langle\psi_j^3\rangle=0,~~~j=1,2
    \label{eq:A11}
\end{equation}
Substituting the averaged expressions from Eq.~\eqref{eq:A11} into Eqs.~\eqref{eq:A3} and \eqref{eq:A4}, we recover the effective dynamics of the primary-oscillator--NES system, as given by Eqs.~\eqref{eq:effective_dyn_PO} and \eqref{eq:effective_dyn_NES}.

\begin{equation}
    \begin{split}
        \ddot{X_1}&+\gamma_1\left(X_1^2+\dfrac{g^2}{2\Omega^4}-1\right)\dot{X_1}+\left(\omega_0^2+\dfrac{3\alpha_1g^2}{2\Omega^2}\left(\dfrac{1}{\Omega^2+\gamma_2^2}\right)\right)X_1+\gamma_2\dot{X}_2+\alpha_1 X_1^3\\
        &+\dfrac{3\alpha_2\epsilon^2g^2}{2\Omega^2}\left(\dfrac{1}{\epsilon^2\Omega^2+\gamma_2^2}\right) X_2+\alpha_2X_2^3=0
    \end{split}
    \label{eq:A12}
\end{equation}

and
\begin{equation}
    \begin{split}
        \epsilon\ddot{X}_2+\gamma_2\dot{X}_2+\dfrac{3\alpha_2\epsilon^2g^2}{2\Omega^2}\left(\dfrac{1}{\epsilon^2\Omega^2+\gamma_2^2}\right) X_2+\alpha_2X_2^3=\epsilon\ddot{X}_1
    \end{split}
    \label{eq:A13}
\end{equation}

%\appendix

\section{Signal-level \texorpdfstring{$Q$}{Q}-estimation from a finite record}
\label{app:qfactor}

We estimate spectrally focused $Q$--factors from uniformly sampled time records of the primary oscillator
displacement $x_1(t)$ and of the NES relative motion $x_1(t)-x_2(t)$ over an observation
window of duration $T$. For a prescribed carrier frequency $\delta$, we form the complex
projections
\begin{align}
\hat Y(\delta)
  &= \frac{2}{T}\int_{0}^{T} x_1(t)\,e^{-i\delta t}\,\mathrm{d}t,
  \label{eq:b1}\\[4pt]
\widehat{Y_{\mathrm{rel}}}(\delta)
  &= \frac{2}{T}\int_{0}^{T} \bigl(x_1(t)-x_2(t)\bigr)\,e^{-i\delta t}\,\mathrm{d}t,
  \label{eq:b2}
\end{align}
computed numerically via the composite trapezoid rule (equivalently, the DFT bin nearest
$\delta$). The resulting metrics are
\begin{align}
Q_{\mathrm{PO}}
  &= \frac{|\hat Y(\delta)|}{\delta^2\,\max_{t}|x_1(t)-x_2(t)|+\varepsilon}
   = \frac{2}{T}\,
      \frac{\sqrt{\Bigl(\displaystyle\int_{0}^{T} x_1(t)\cos(\delta t)\,\mathrm{d}t\Bigr)^{2}
                 +\Bigl(\displaystyle\int_{0}^{T} x_1(t)\sin(\delta t)\,\mathrm{d}t\Bigr)^{2}}}
           {\delta^2\,\max_{t}|x_1(t)-x_2(t)|+\varepsilon},
  \label{eq:qpo_app}\\[8pt]
Q_{\mathrm{NES}}
  &= \frac{|\widehat{Y_{\mathrm{rel}}}(\delta)|}{\delta^2\,\max_{t}|x_1(t)|+\varepsilon}
   = \frac{2}{T}\,
      \frac{\sqrt{\Bigl(\displaystyle\int_{0}^{T} \bigl(x_1(t)-x_2(t)\bigr)\cos(\delta t)\,\mathrm{d}t\Bigr)^{2}
                 +\Bigl(\displaystyle\int_{0}^{T} \bigl(x_1(t)-x_2(t)\bigr)\sin(\delta t)\,\mathrm{d}t\Bigr)^{2}}}
           {\delta^2\,\max_{t}|x_1(t)|+\varepsilon},
  \label{eq:qnes_app}
\end{align}
with $\varepsilon=10^{-12}$ to avoid division by zero, and $\max_{t}$ denoting the maximum
over $t\in[0,T]$.

\paragraph{Normalization rationale.}
In $Q_{\mathrm{PO}}$, the denominator plays the role of an effective forcing amplitude
exerted by the NES on the primary oscillator. Indeed, if at a given coupling $g$ the NES relative motion
is approximated at the carrier frequency as
\[
x_1(t)-x_2(t) \approx A_{\mathrm{rel}}\,e^{-i\delta t},
\qquad
A_{\mathrm{rel}}=\max_{t}|x_1(t)-x_2(t)|,
\]
then the associated acceleration--type forcing acting on the primary oscillator scales as
\[
\ddot{x}_{\mathrm{rel}}(t)\sim -\delta^{2}A_{\mathrm{rel}}\,e^{-i\delta t},
\]
so that the effective forcing amplitude is proportional to $\delta^{2}\max_{t}|x_1(t)-x_2(t)|$.
We retain this normalization in analogy with conventional vibrational--resonance algorithms
\cite{landa2000vibrational}, where the $Q$--factor is defined as the Fourier--projected
response amplitude divided by the amplitude of the slow forcing.

In analogy with $Q_{\mathrm{PO}}$, the denominator in $Q_{\mathrm{NES}}$ represents the
effective slow forcing amplitude exerted by the primary oscillator on the NES: if
$x_1(t)\approx A_1 e^{-i\delta t}$ with $A_1=\max_t|x_1(t)|$, then the corresponding
acceleration--type forcing scale is $\delta^{2}A_1$, which motivates the normalization.

\bibliographystyle{apsrev4-1}
\bibliography{new_refs_vr_internal}

\end{document}